\documentclass[aps,twocolumn,superscriptaddress,groupedaddress]{revtex4}  
\usepackage{graphicx}  
\usepackage{dcolumn}   
\usepackage{bm}        
\usepackage{amssymb}   
\usepackage{amsthm}
\usepackage{amsmath}
\usepackage{latexsym}
\usepackage{lscape}
\usepackage{epsfig}
\usepackage{pstricks}
\usepackage{amsfonts}
\usepackage{exscale}
\usepackage{relsize}

\begin{document}


\author{Duokui Yan \footnote{duokuiyan@buaa.edu.cn}} \affiliation{School of Mathematics and System Science, Beihang University, Beijing 100191, China}

\author{Tiancheng Ouyang \footnote{ouyang@math.byu.edu}}  \affiliation{Department of Mathematics, Brigham Young University, Provo, UT 84602, USA}

\title{New phenomenons in the spatial isosceles three-body problem} 


\begin{abstract}
In this work, we study the periodic orbits in the spatial isosceles three-body problem. These periodic orbits form a one-parameter set with a rotation angle $\theta$ as the parameter. Some new phenomenons are discovered by applying our numerical method. The periodic orbit coincides with the planar Euler orbit when $0 < \theta \leq 0.32 \pi$ and it changes to a spatial orbit when $0.33 \pi \leq \theta < \pi$. Eventually, the spatial orbit becomes a planar collision orbit when $\theta=\pi$. Furthermore, an oscillated behavior is found when $\theta=\pi/2$, which is chaotic but bounded under a small perturbation. As another application of our numerical method, 7 new periodic orbits are presented in the end.
\end{abstract}

\maketitle
%


\section{Introduction}
 The Newtonian N-body problem studies the motions of N gravitationally interacting point masses, which is easy to define but difficult to solve for $N \geq 3$.  ``Despite efforts by outstanding mathematicians for over 200 years, the problem remains unsolved to this day." \cite{SM} However, the study of the N-body problem has brought an abundance of original ideas which severed as a stimulus for several branches of mathematics. Meanwhile, new theories and methods are introduced to the N-body problem, which lead to the development in this area.
 
Application of variational method to periodic orbits is one of the recent breakthroughs in the N-body problem. Actually, back to 1890's, Poincar\'{e} \cite{poincare} have tried so, but he didn't  succeed because of two major difficulties. One is the lack of coercivity due to the vanishing at infinity of the force fields. The other is the possible existence of collision: the Lagrangian action stays finite even when some of the bodies are colliding. Until recently in 2000, Chenciner and Montgomery \cite{CM} considered the action minimizer over a suitable symmetric loop space and successfully conquered the two difficulties. They showed the existence of the famous figure-eight orbit in the planar three-body problem, which was discovered by Moore \cite{MO} in 1993. Following their ideas of imposing symmetry constraint, many new periodic orbits have been discovered and proved rigorously. 

Besides the figure-eight orbit \cite{MO, CM}, there are not so many known nontrivial orbits in the three-body problem. Recently, \v{S}uvakov and Dmitra\v{s}inovi\'{c} \cite{SD} discovered 13 distinct orbits, which is quite a surprise. These new orbits have rich solution structures and their result convinces us that there exists numerous complicated periodic orbits in the planar three-body problem.  For the spatial three-body problem, very few periodic orbits have been found so far. The set of periodic orbits in the spatial isosceles three-body problem is one of the known sets in 3D. It is actually a one-parameter set, with rotation angle $\theta$ as the parameter. The existence of these orbits has been studied in several works. In 2004, Offin \cite{OF, MHO} claimed an existence proof in a symmetric subspace $\Gamma$ by a global variational method, where $\Gamma$ is
\begin{eqnarray*}
\Gamma = \bigg\{ (q_1, q_2, q_3)  \big|  \sum_{i=1}^3 m_iq_i=0,  q_{2x}=  q_{2y}=0,  q_{1z}=q_{3z}  \bigg\},
\end{eqnarray*}
and $q_i=(q_{ix}, q_{iy}, q_{iz})$ is the coordinate of the mass $m_i \ (i=1,2,3)$. Later in 2009, Shibayama \cite{SH} provided a different variational proof. However, there are still lots of questions concerning the properties of this set of orbits. For example, is this set of orbits always spatial? If not, when will the periodic orbit be planar?  For different rotation angle $\theta$, what does the orbit look like? Does there exist any strange motion in this set of orbits? Is there a stable orbit in this set?

In this work, we apply our variational method and search for all possible motions of the periodic orbits in the equal-mass spatial isosceles three-body problem. A global picture of this set of orbits are presented for the first time. When $\theta \in (0,  0.32 \pi]$, the orbit is actually the planar Euler orbit. When $\theta \in [0.33\pi,  \pi)$, the orbit is 3D and it looks quite simple for special angles, such as $\theta=\pi/2,  3 \pi/4$, etc. In particular, when $\theta=\pi/2$, an oscillated behavior is located as we perturb the initial condition. 

In the end, we present 7 new periodic orbits in the N-body problem as another application of our method. Initial conditions are included for convenience.   

\section{Numerical method}
Variational method is an important tool when studying periodic orbits in the N-body problem. A standard approach can be found in \cite{Van}, where he introduced the Fourier series and minimized the Lagrangian action over a whole period of an orbit. Many periodic orbits were found by this method. However, it is not so powerful when studying the detailed information of a specific set of periodic orbits. In our variational method,  instead of searching a whole period of an orbit or imposing symmetry constraints, we concentrate on only one part of a desired periodic orbit in a fixed time interval $[0,1]$. To make this method clear, we describe it in two steps.

\begin{figure}[htb]
\includegraphics[scale=0.28]{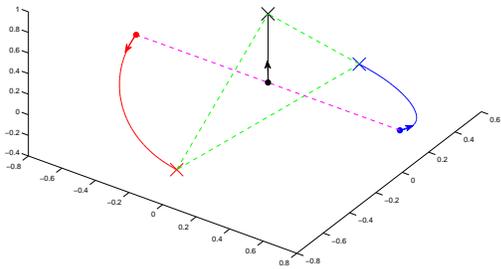}
\caption{Demonstration of one part of a spatial isosceles periodic orbit: from a collinear configuration (in dots) to an isosceles configuration (in crosses) when the middle body reaches its maximum height.}
\label{fig:describe1}
\end{figure}

\begin{figure}[htb]
\centering
\begin{center}
\psset{xunit=1in,yunit=1in}
\begin{pspicture}(-2.5, 0.7)(-0.5,1.2)
\psline[linestyle=dashed, linewidth=0.2pt](-2,1)(-0.8,1)
\psdots[dotsize=4pt 1](-2,  1)(-1.4, 1)(-0.8, 1)
\rput(-2.5, 1){$Q(0):$}
\rput(-1.4, 1.12){2}
\rput(-2.05 ,0.93 ){1}
\rput( -0.74, 0.93){3}
\end{pspicture}
\end{center}
\begin{center}
\psset{xunit=1in,yunit=1in}
\begin{pspicture}(-0.1, 0.45)(1.8,1.85)
\psdots[dotsize=2pt 1](0.8, 1)
\rput(0.86, 1.05){0}
\psline[linestyle=dashed, linewidth=0.2pt](0.8, 1)(0.8, 1.6)
\rput(0.8, 1.6){x}
\rput(0.8, 1.74){2}
\psline[linestyle=dashed, linewidth=0.2pt](0.8, 1)(0.8, 0.7)
\rput( 0.35, 0.54){x}
\rput(1.24, 0.86){x}
\psline[linestyle=dashed, linewidth=0.2pt]( 0.35, 0.54)(1.25 , 0.86)
\psline[linestyle=dashed, linewidth=0.2pt]( 0.8, 1.6)(0.35, 0.54)
\psline[linestyle=dashed, linewidth=0.2pt]( 0.8, 1.6)(1.24, 0.86)
\rput( 0.25, 0.5){1}
\rput( 1.36,0.9 ){3}
\rput(-0.1 ,1){$Q(1):$}
\psline[linestyle=dashed, linewidth=0.2pt](0.4 ,0.7)( 1.35,0.7)
\pscurve[linewidth=0.2pt]{<->}(1, 0.7)(1.06, 0.75)(1.1,  0.8)
\psline[linewidth=0.5pt]{<-}(1.16 ,0.75)(1.65 ,0.75)
\rput(1.75 , 0.73){$\theta$}
\end{pspicture}
\end{center}
\caption{$Q(0)$ and $Q(1)$ in a spatial isosceles periodic orbit.}
\label{fig:describe0}
\end{figure}
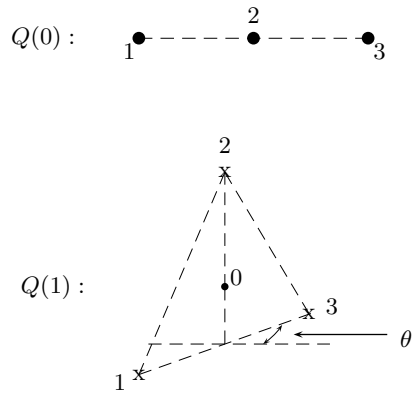

The first step is to choose boundary configurations $Q(0)$ and $Q(1)$ in $[0,1]$, where $Q(0)$ and $Q(1)$ are both $N \times 2$ (planar orbits) or $N \times 3$ matrices (spatial orbits). $Q(0)$ is the position matrix at $t=0$ and $Q(1)$ is the one at $t=1$. Each row of the matrices represents the coordinate of each body at the corresponding time. For instance, the first row of $Q(0)$ is the position vector of the first body at $t=0$, and the first row of $Q(1)$ is its position at $t=1$, and so on. In general, the number of free variables in $Q(0)$ and $Q(1)$ are $4N$ (for planar orbits) or $6N$ (for spatial orbits). For $N \geq 3$, this number is always big. It is better to decrease it a lot so that numerical programs can handle these variables efficiently. By setting the center of mass to be 0, we can lower the number of variables a little bit. Furthermore, the boundary configurations $Q(0)$ and $Q(1)$ are often in special shapes, which greatly reduce the number of variables. For example, we consider a periodic orbit in the equal-mass spatial isosceles three-body problem, which is called as a spatial isosceles periodic orbit. In each period of such an orbit, one body moves up and down along a vertical line, and the other two bodies rotate about this line. A demonstration of one part of this orbit can be found in Fig.~\ref{fig:describe1}. If one wants to construct it by our method, the two boundary configurations $Q(0)$ and $Q(1)$ can be set as in Fig.~\ref{fig:describe0}: $Q(0)$ is a collinear configuration; $Q(1)$ is an isosceles configuration.

It is natural to define 
\begin{eqnarray}\label{spa0} 
Q(0)=  \begin{bmatrix}
-a   &  0  & 0    \\
0  &  0  &   0   \\
a   & 0   &   0
\end{bmatrix},
\end{eqnarray}
\begin{eqnarray}\label{spa1}
Q(1)= \begin{bmatrix}
-b   &  0  &  -c  \\
0  &  0   &   2c \\ 
b   &   0  &    -c
\end{bmatrix}\begin{bmatrix}
 \cos \theta   &  \sin \theta  &  0   \\
 - \sin \theta   &    \cos \theta    &   0 \\
0     &  0     &    1
\end{bmatrix}.  
\end{eqnarray}
In other words, at $t=0$, the coordinates of the three bodies are $(-a, 0,   0)$, $(0, 0,  0)$ and $(a,  0,   0)$ respectively. At $t=1$,  their positions change to $(-b \cos \theta,  -b \sin \theta,  -c)$,  $(0,  0,   2c)  $ and $(b \cos \theta,  b \sin \theta,  -c)$ correspondingly. 

After the selection of $Q(0)$ and $Q(1)$, a boundary value problem is studied in order to find a solution connecting the two configurations. We introduce the standard Sobolev space $H^{1}([0,1], \chi)$, where $\chi=  \{ q=(q_1^T, q_2^T, \dots, q_N^T)^T \big|  \sum_{i=1}^N  m_{i}q_{i}= 0 \}$ and $q_i  \in \mathbb{R}^2$ or $\mathbb{R}^3$ is a row vector representing the position of mass $m_i$. The Lagrangian action functional $\mathcal{A}: H^1([0, 1], \chi) \mapsto  \mathbb{R}$  is 
\begin{eqnarray}\label{actionformula} 
 \displaystyle \mathcal{A} & =&\int_0^1 L(q, \dot{q}) dt =  \int_0^1 \left( K+U \right) dt,
\end{eqnarray}
 where
\[K= \sum_{i=1}^N \frac{1}{2} m_i |\dot{q}_i|^2, \quad U=\sum_{1\leq i<j \leq N}  \frac{m_i m_j}{ \mid q_i- q_j  \mid } . \] It is known that, for two different configurations: $Q(0)$ and $Q(1)$, a minimizing path $\mathcal{P}$ connecting them can be generated as follows,
\begin{eqnarray*}
\mathcal{A} (\mathcal{P})& = &\inf_{\{q(0)=Q(0), \,  q(1)=Q(1), \, q \in \chi \} } \mathcal{A},
\end{eqnarray*}
 which is a solution of the N-body problem. Numerically, a standard way to create the path $\mathcal{P}$ is the finite difference method. To speed up the process, one can generate a rough path with a small partition number at the beginning, for example, $100$. Based on this rough path, another $100$ points can be interpolated to make it better. Keep repeating the interpolation a few times, one can end up with a path which has a sufficient amount of points. In our numerical search, the final number of points in a path is usually set as $1,000\times 2^5=32,000$. At the same time, the Lagrangian action of this path $ \mathcal{A}(\mathcal{P})= \int_0^1 L (q, \dot{q}) \, dt= \int_0^1 (K+U) \, dt $ is calculated by a Riemann sum. 

The second step is to free several parameters in $Q(0)$ and $Q(1)$, and minimize the Lagrangian action $ \mathcal{A}=\int_0^1 L(q, \dot{q}) \, dt $ in \eqref{actionformula} over these parameters. For given values of the parameters, $Q(0)$ and $Q(1)$ are fixed. A path connecting them can be generated by the first step.  Correspondingly, its Lagrangian action $\mathcal{A}$ can be calculated.  The local minimization of $\mathcal{A}$ over these parameters is realized by a Matlab program. A first variation argument shows that local minimizers of $\mathcal{A}$  must be solutions of the N-body problem. If one chooses the right parameters in $Q(0)$ and $Q(1)$, a local minimizer can be one part of a periodic orbit. For example, in the setting of $Q(0)$ (as in  Eqn.~\eqref{spa0}) and $Q(1)$ (as in  Eqn.~\eqref{spa1}), there are four variables: $a$, $b$, $c$ and $\theta$. To find a spatial isosceles orbit (Fig.~\ref{fig:describe1}), one needs to set $a$, $b$, $c$ as parameters and minimize the Lagrangian action $\mathcal{A}$ over them. The variable $\theta$ in $Q(1)$ is regarded as a fixed angle. It is clear that the period of this orbit is not $1$. In fact, one can generate the periodic orbit by reflecting and rotating this minimizer several times.  

\section{Application to spatial isosceles orbits}
\begin{figure}[htb]
\includegraphics[scale=0.8]{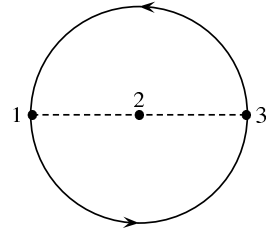}
\caption{\label{fig:euler} Circular Euler orbit.}
\end{figure}

In this section, we apply our variational method to the spatial isosceles orbits in the equal-mass three-body problem. In this case, $N=3$ and $m_1=m_2=m_3=1$. The boundary configurations are defined as follows
\begin{eqnarray}\label{Q00}
Q(0)= \begin{bmatrix}
-a   &  0  & 0    \\
0  &  0  &   0   \\
a   & 0   &   0
\end{bmatrix},
\end{eqnarray}
\begin{eqnarray}\label{Q11}
Q(1)= \begin{bmatrix}
-b   &  0  &  -c  \\
0  &  0   &   2c \\ 
b   &   0  &    -c
\end{bmatrix}\begin{bmatrix}
 \cos \theta   &  \sin \theta  &  0   \\
 - \sin \theta   &    \cos \theta    &   0 \\
0     &  0     &    1
\end{bmatrix},  
\end{eqnarray}
where $a, b, c > 0$. Path $\mathcal{P}_0$ is the Lagrangian action minimizer and it satisfies
\begin{eqnarray}\label{minisosceles}
& & \mathcal{A}(\mathcal{P}_0) \\
&=& \min_{ \{ a, b ,c >0  \}  \ }  \inf_{\{ q(0)=Q(0), \, q(1)=Q(1)\} }  \int_0^1L(q, \dot{q}) \, dt. \nonumber 
\end{eqnarray}
As in Fig.~\ref{fig:describe1} and Fig.~\ref{fig:describe0}, we set body 2 to be the one moving up and down on the vertical line ($z-$axis). The other two bodies are named as bodies 1 and 3. The rotation angle $\theta$ measures how much bodies 1 and 3 rotate on the $xy$ plane during the time interval $[0, 1]$.  For any given $\theta \in (0, \pi]$, there exists a minimizing path $\mathcal{P}_0$ which is a part of a periodic or quasi-periodic orbit. We search for the initial conditions of the periodic orbits when $\theta$ increases from $0$ to $\pi$ with a step $0.01 \pi$. The motions of the periodic orbits are checked by simulators. There are basically four different types of motions. We interpret them in detail case by case.\\\\
\textbf{(I):} The periodic orbit is the planar Euler orbit when $\theta \in (0, 0.32 \pi]$. A picture of a circular Euler orbit is given in Fig.~\ref{fig:euler}. In the graph, body 2 always stays at origin and bodies 1 and 3 run on a circular orbit. 

\begin{figure}[htb]
\includegraphics[scale=0.25]{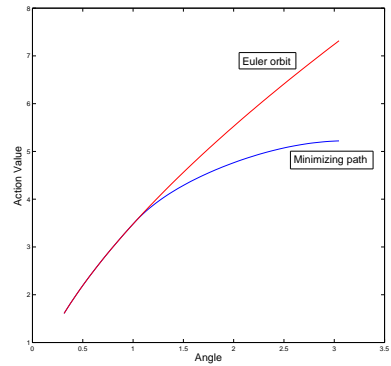}
\caption{\label{fig:compare} Lagrangian actions of the Euler orbit $\mathcal{P}_e$ and the minimizing path $\mathcal{P}_0$ with respect to each rotation angle $\theta$.}
\end{figure}

We denote the piece of an Euler orbit by $\mathcal{P}_e$, which is a local action minimizer connecting $Q(0)$ (as in Eqn.\eqref{Q00}) and $Q(1)$ (as in Eqn.\eqref{Q11}) in the time interval $[0, 1]$. Actually for any $\theta \in (0, \pi]$, there exists a local action minimizer $\mathcal{P}_e$, which is a part of an Euler orbit. As $\theta$ increases, its Lagrangian action $\mathcal{A}(\mathcal{P}_e)$ may not always be the absolute minimum. We compare the Lagrangian actions of the Euler orbit $\mathcal{P}_e$ and the minimizing path $\mathcal{P}_0$ in Fig.~\ref{fig:compare}.
In this graph, the horizontal axis is the rotation angle $\theta$ and the vertical axis is the Lagrangian action $\mathcal{A}$ of a path in $[0, 1]$. The two curves  in Fig.~\ref{fig:compare} are about to be different when the rotation angle $\theta$ is around $1$. A closer look shows that it is between $0.32 \pi= 1.0053$ and $0.33 \pi= 1.0367$. 

In fact, it can be illustrated in the following way. When $\theta$ is small, the average angular velocity is not big enough to pull body 2 up. As $\theta$ increases, there may exist two local action minimizers: the Euler orbit and the spatial isosceles orbit. Fig.~\ref{fig:compare} shows the spatial isosceles orbit has a smaller action when $\theta$ is relatively big. We call $\theta_0$ a critical rotation angle when the corresponding action minimizer $\mathcal{P}_0$ changes from an Euler orbit to a spatial isosceles orbit. In our case, $\theta_0 \approx 0.32 \pi$. Generally speaking, if one studies the set of spatial isosceles orbits with masses $[1,\,  m, \, 1]$, it is reasonable to expect a critical angle $\theta_0(m)$ for each mass $m$ of body 2, where $\theta_0(1)= \theta_0$. \\\\
\textbf{(II):} For $\theta \in [0.33 \pi, \pi)$, the minimizing path $\mathcal{P}_0$ (in Eqn.~\eqref{minisosceles})  is a 3D periodic orbit. Here we provide the initial conditions of the spatial isosceles periodic orbits corresponding to several rotation angles $\theta$ in Table \ref{tab:table2}.
\begin{table}
\caption{\label{tab:table2} This table contains the values of $a$, $v_y$ and $v_z$ in the initial condition matrix (see Footnote $a$.) for the spatial isosceles periodic orbits with respect to the rotation angles $\theta$. }
\begin{ruledtabular}
\begin{tabular}{cccccccc}
 $\theta$ & $a$  & $v_y$   &$v_z$&
  $\theta$ & $a$ &$v_y $ &$v_z$ \\
\hline\\
$\pi/2 $ & 0.7453 &  0.7335 & 0.6733  & $ 4\pi/5  $   & 0.7687 & 0.2950 & 0.7005 \\\\
$\pi/3$  & 1.0253 & 1.0500 & 0.2159 & $ 3\pi/8$ & 0.9039 & 0.9464 & 0.4436  \\\\
$ 2\pi/3$ & 0.7182 & 0.4970 & 0.7354 & $  5\pi/8 $& 0.7133 & 0.5568 & 0.7340 \\\\
$ 3 \pi/4 $ & 0.7451 & 0.3723 & 0.7188 &  $ 7 \pi/8$ & 0.8057 & 0.1803 &  0.6703  \\\\
$ 2\pi/5 $ & 0.8542 & 0.8959 & 0.5171 &  $ 7\pi/10$ & 0.7265  & 0.4479 & 0.7315 \\\\
$ 3\pi/5$  & 0.7137 & 0.5921 & 0.7292 & $ 9\pi/10$  & 0.8158 & 0.1428 &  0.6621 \\
\end{tabular}
\end{ruledtabular}
\footnotetext[1]{The matrix of initial condition has the following form: \\
$ \begin{bmatrix}
q_1 & \dot{q}_1 \\
q_2 & \dot{q}_2 \\
q_3 & \dot{q}_3
\end{bmatrix}=\begin{bmatrix}
-a     &     0     &     0        &     0      &         -v_y       &    -v_z           \\
 0               &    0       &       0      &      0      &         0        &      2 v_z \\
  a   &     0     &         0    &    0        &       v_y            &      -v_z         
  \end{bmatrix}$.}
\end{table}
\begin{figure}[htb]
\begin{minipage}[t]{0.48\linewidth}
\centering
\includegraphics[width=1.8in]{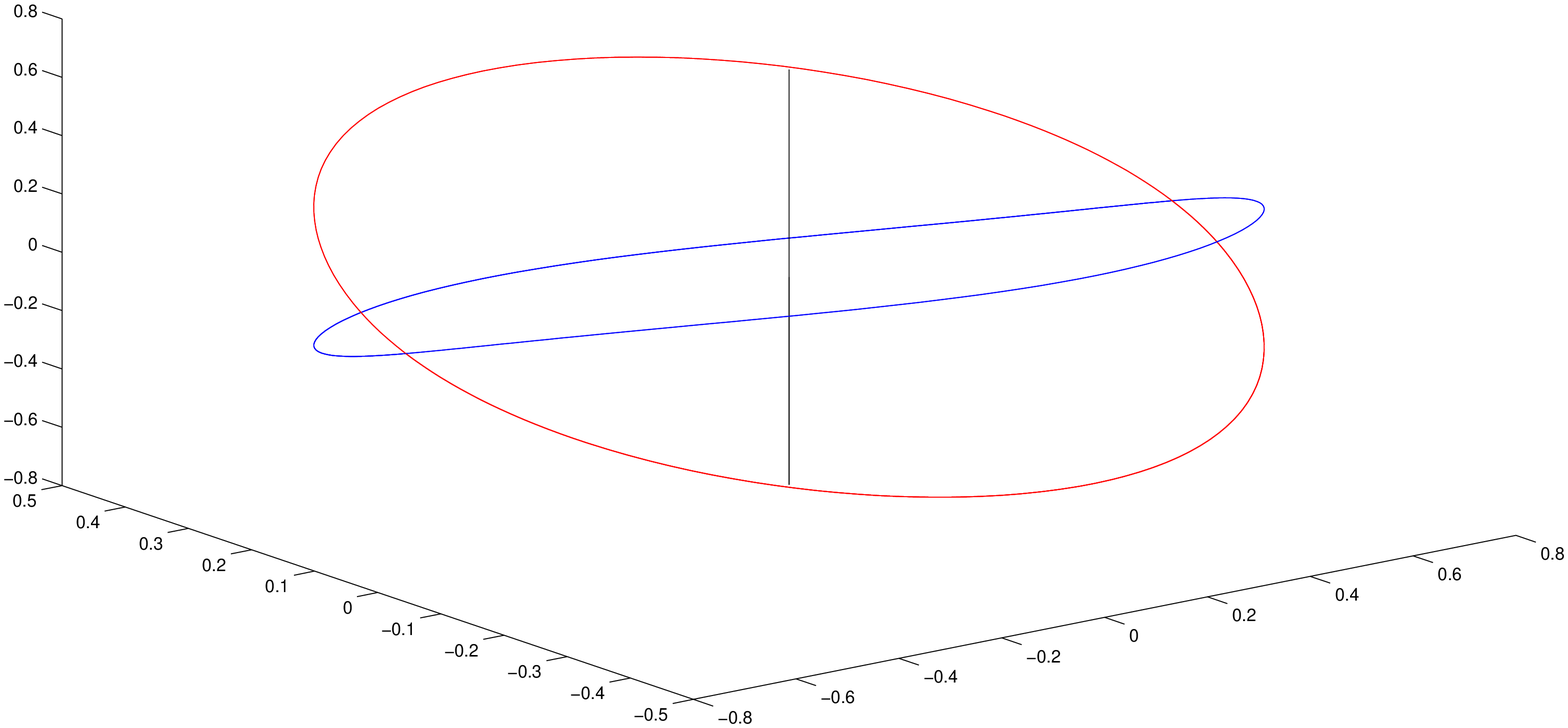}
\caption{$\theta=\pi/2$.}
\label{fig:c1}
\end{minipage}%
\begin{minipage}[t]{0.48\linewidth}
\centering
\includegraphics[width=1.8in]{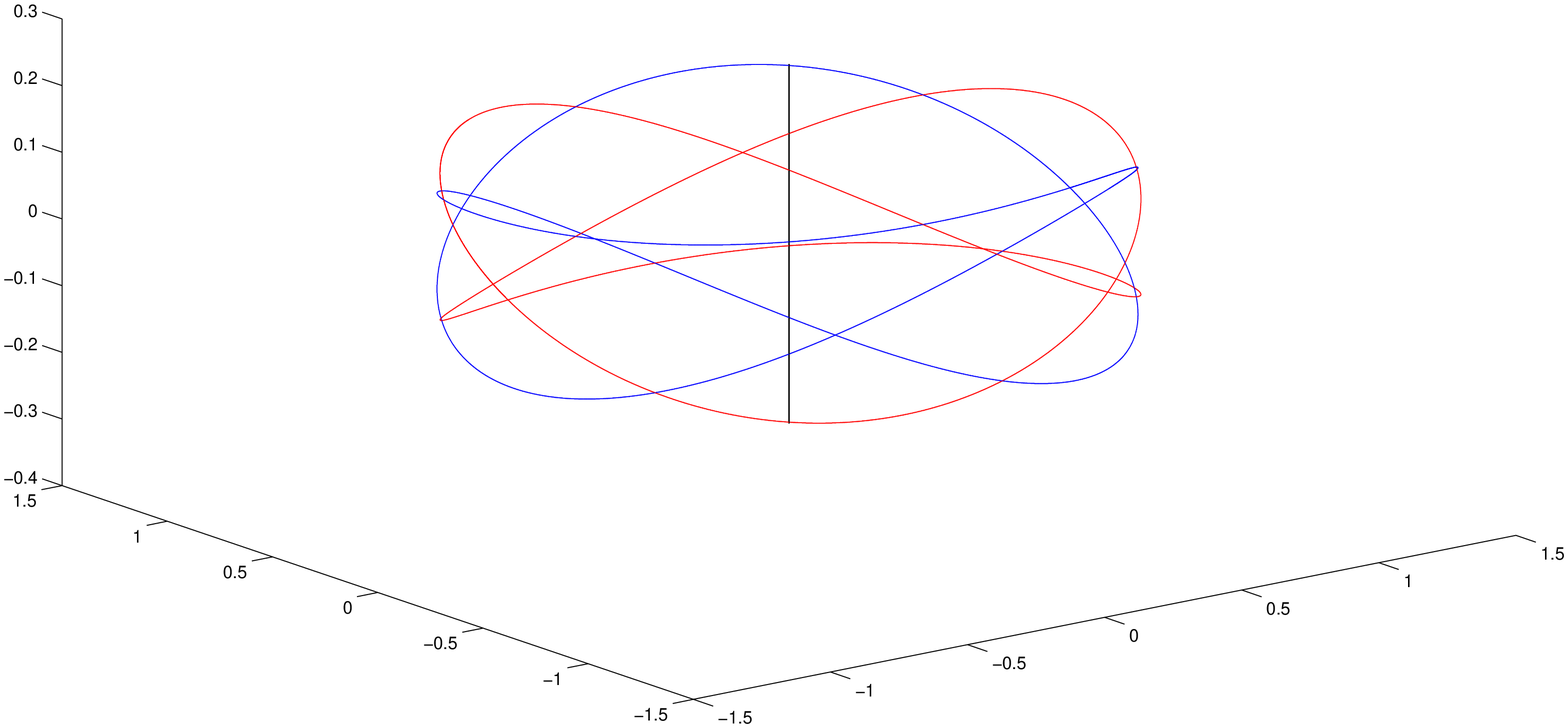}
\caption{$\theta=\pi/3$.}
\label{fig:c2}
\end{minipage}
\begin{minipage}[t]{0.48\linewidth}
\centering
\includegraphics[width=1.8in]{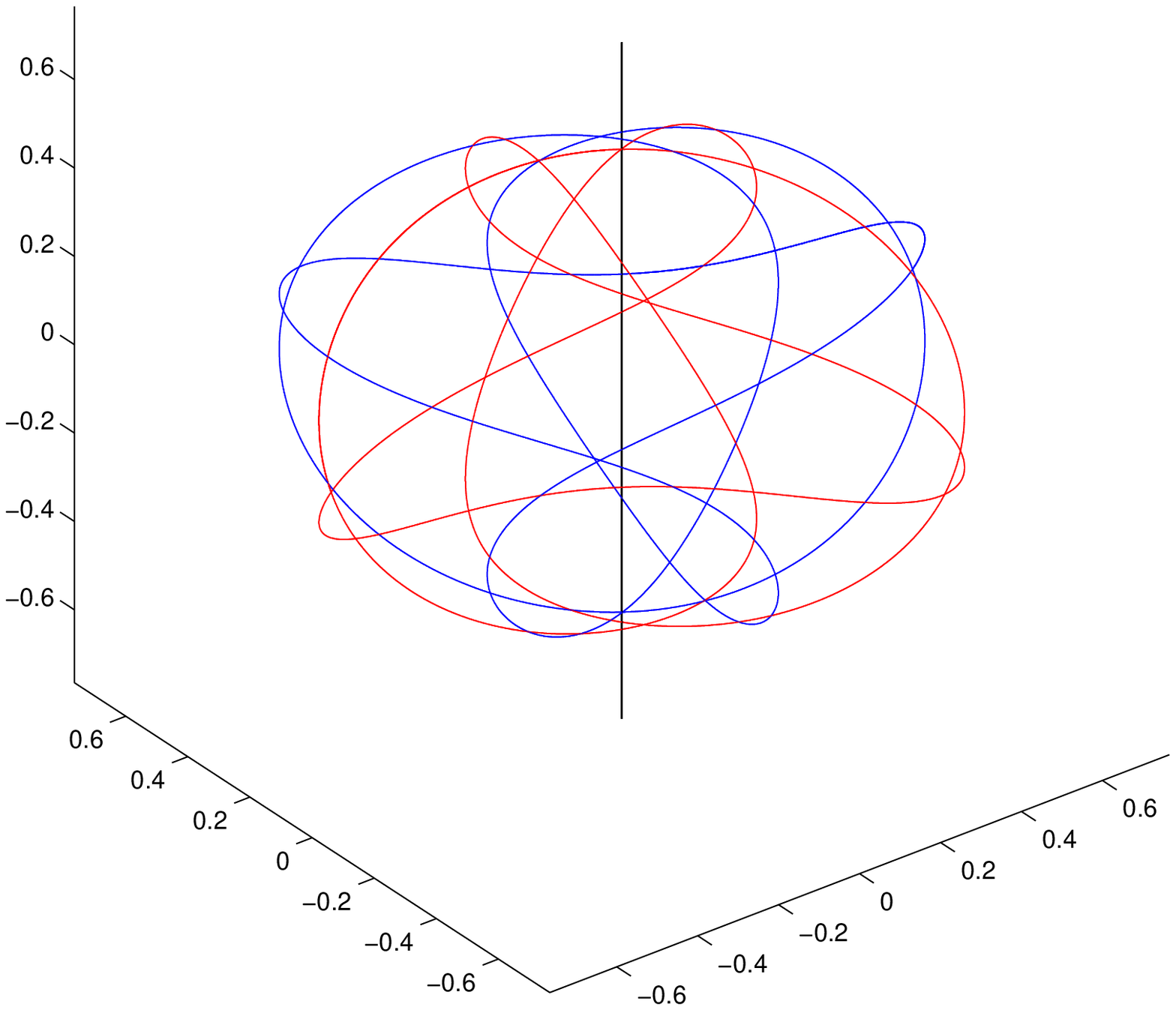}
\caption{$\theta=2\pi/3$.}
\label{fig:c3}
\end{minipage}%
\begin{minipage}[t]{0.48\linewidth}
\centering
\includegraphics[width=1.8in]{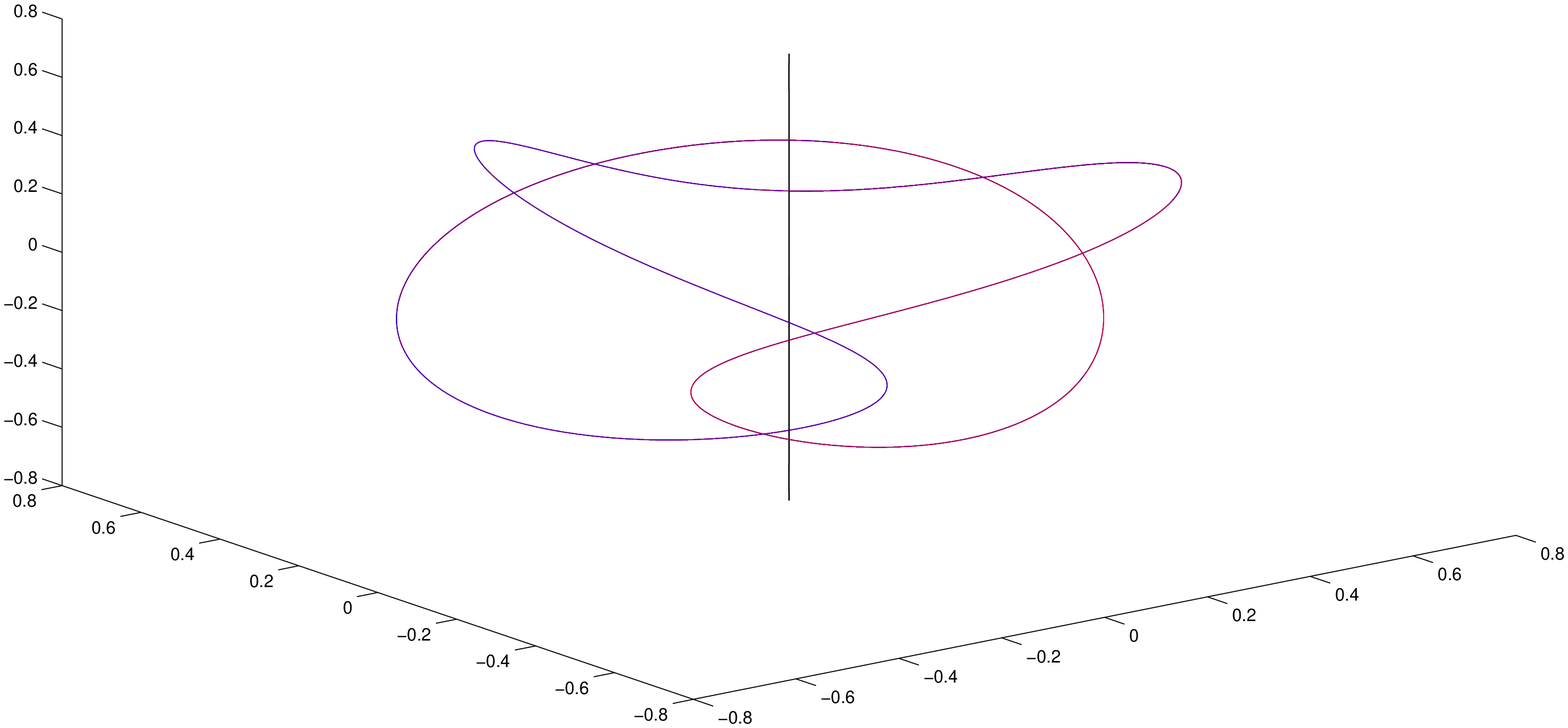}
\caption{$\theta=3\pi/4$.}
\label{fig:c4}
\end{minipage}
\begin{minipage}[t]{0.48\linewidth}
\centering
\includegraphics[width=1.8in]{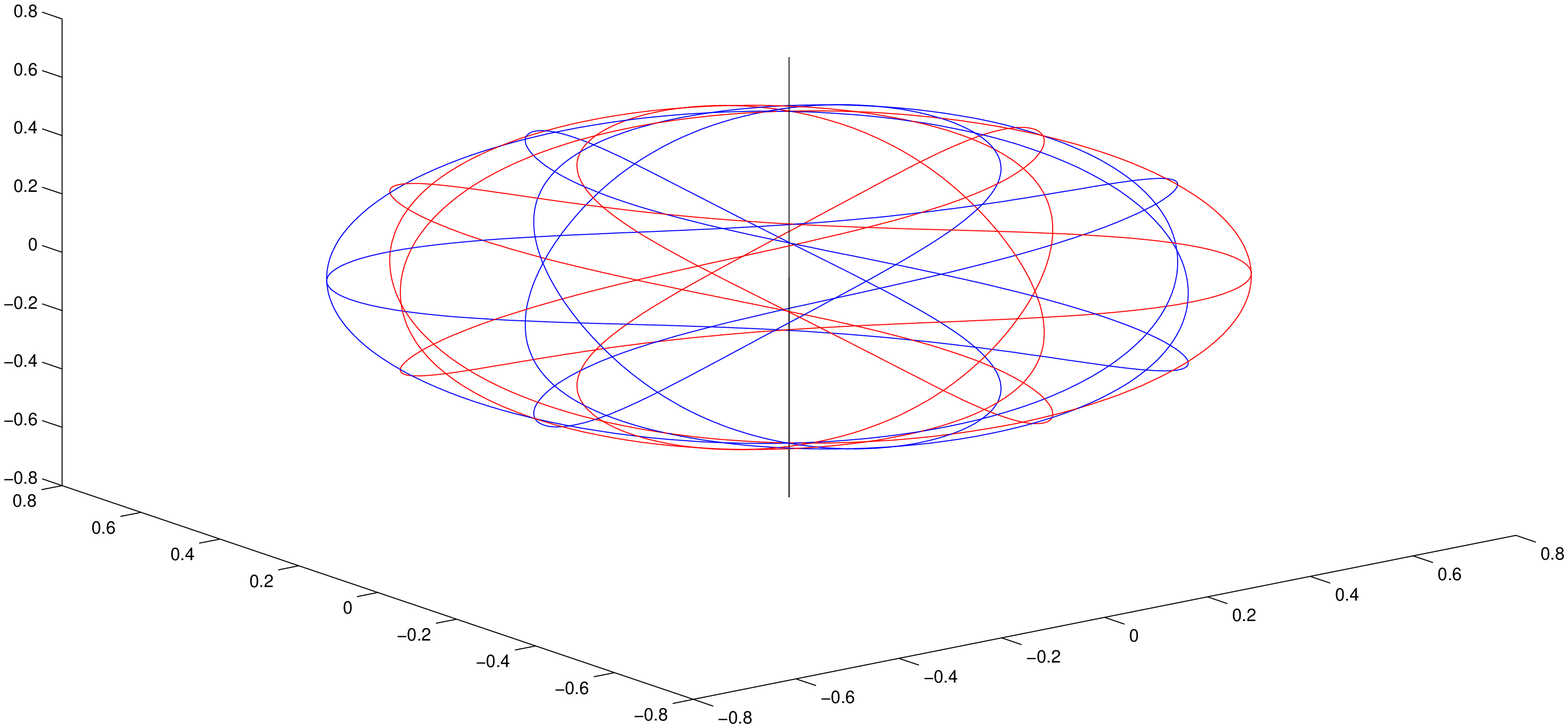}
\caption{$\theta=3\pi/5$.}
\label{fig:c5}
\end{minipage}%
\begin{minipage}[t]{0.48\linewidth}
\centering
\includegraphics[width=1.8in]{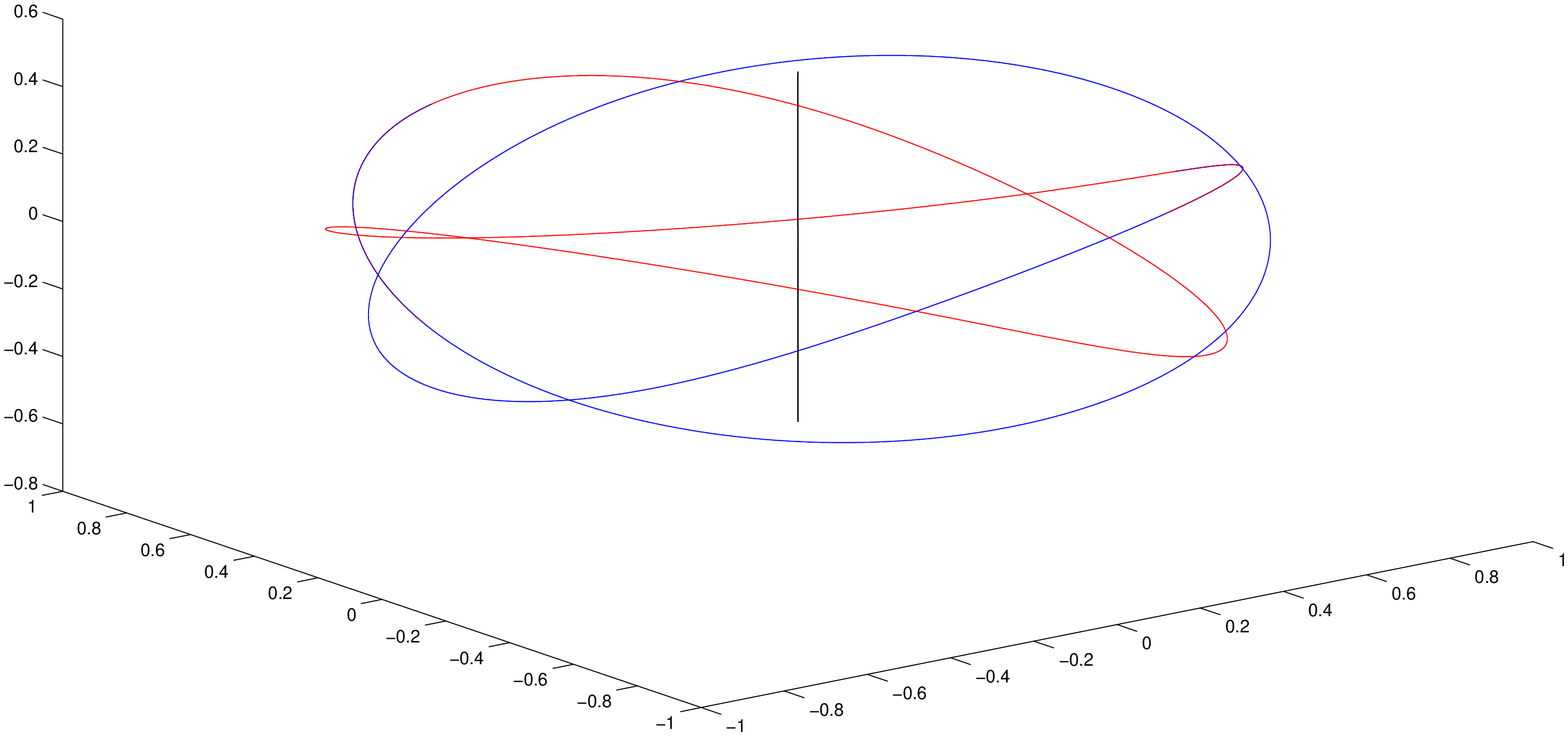}
\caption{$\theta=3\pi/8$.}
\label{fig:c6}
\end{minipage}
\begin{minipage}[t]{0.48\linewidth}
\centering
\includegraphics[width=1.8in]{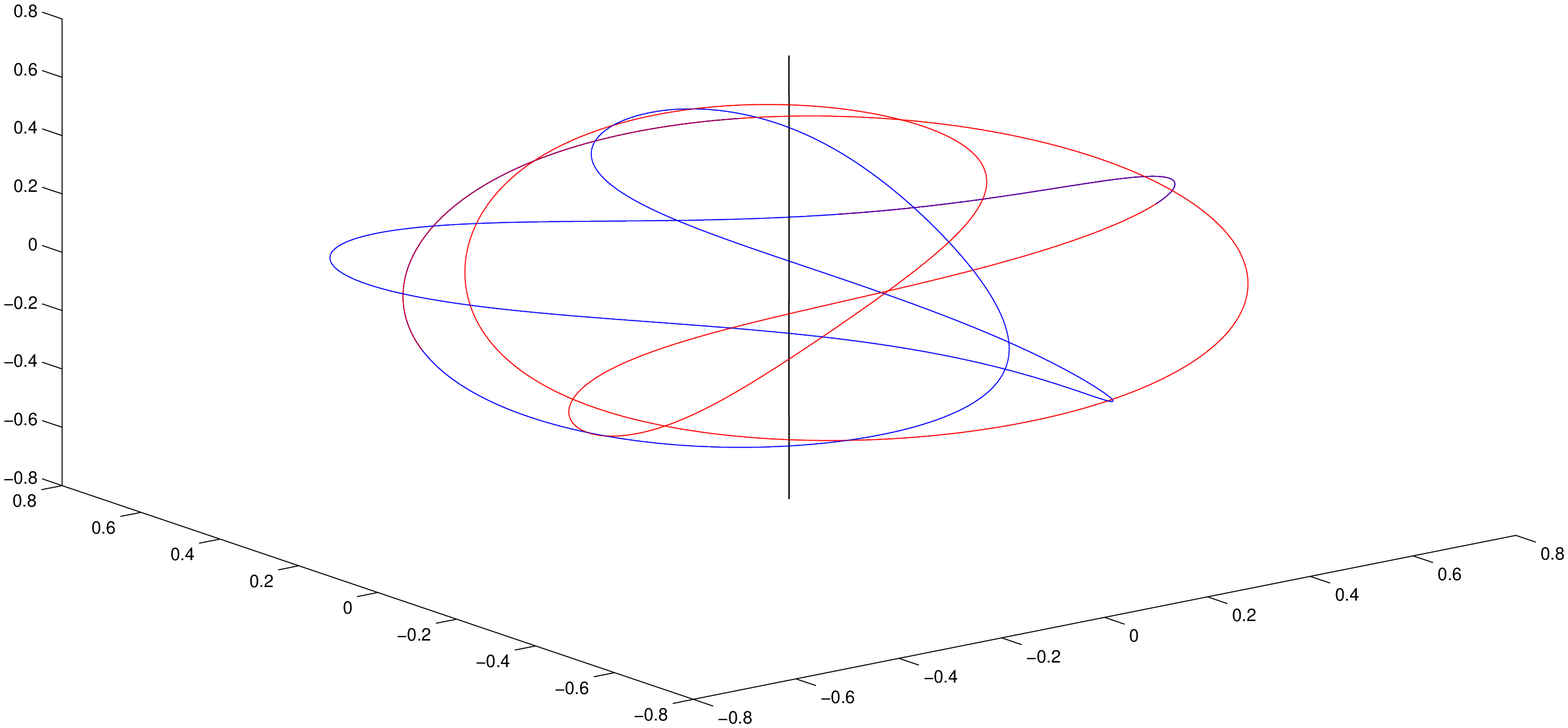}
\caption{$\theta=5\pi/8$.}
\label{fig:c7}
\end{minipage}%
\begin{minipage}[t]{0.48\linewidth}
\centering
\includegraphics[width=1.8in]{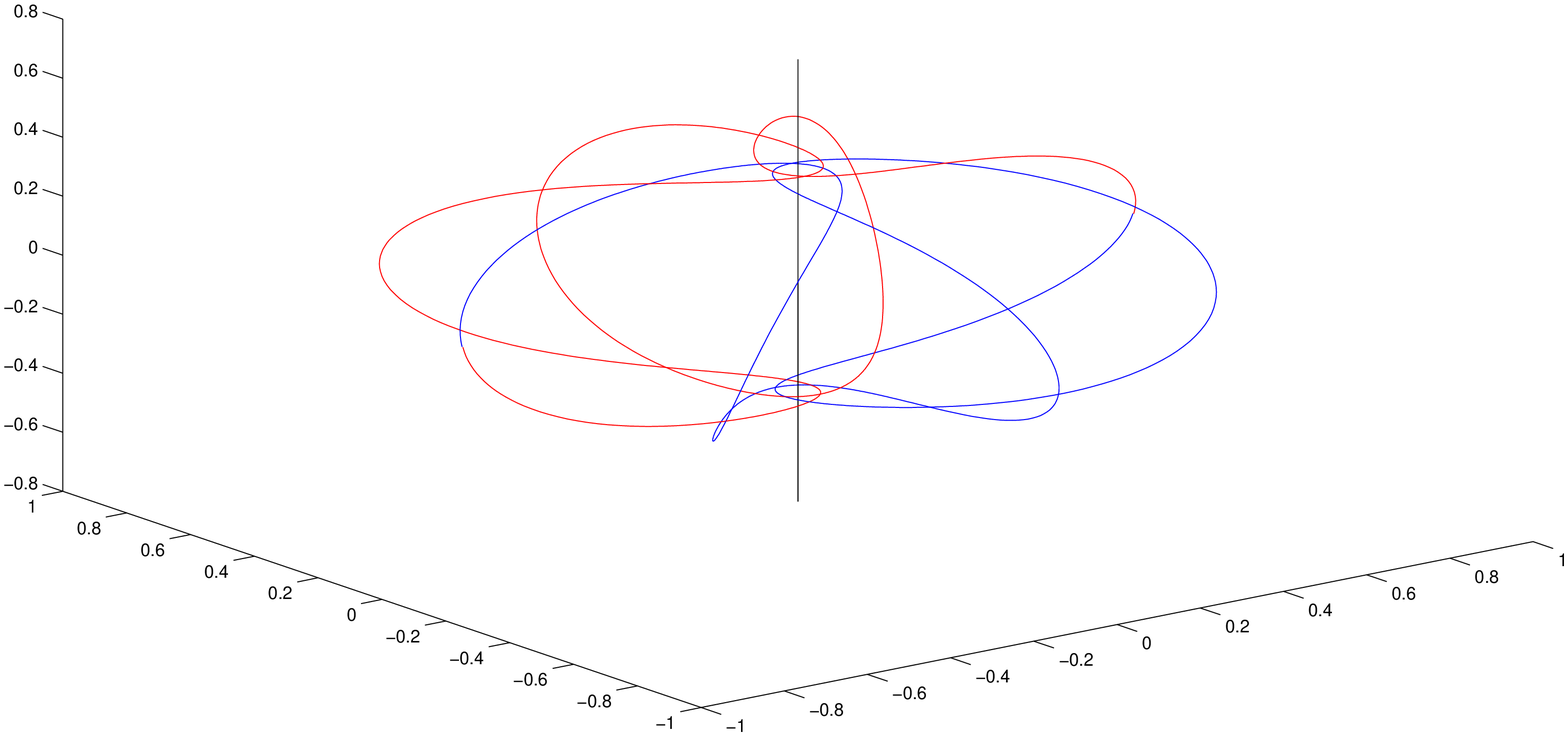}
\caption{$\theta=7\pi/8$.}
\label{fig:c8}
\end{minipage}
\end{figure}
Eight pictures of motions (from Fig.~\ref{fig:c1} to Fig.~\ref{fig:c8}) are  presented here. Note that when $\theta= \pi/2$ or $\theta=3\pi/4$, the picture of motion is quite simple. 

The stability of this set of periodic orbits in the symmetric subspace $\Gamma$ are also checked numerically, where $\Gamma$ is
\begin{eqnarray} \label{gammadef}
\Gamma = \bigg\{ (q_1, q_2, q_3)  \big| \sum_{i=1}^3 q_i=0,  q_{2x}=  q_{2y}=0,  q_{1z}=q_{3z}  \bigg\}.
\end{eqnarray}
It turns out that the spatial isosceles orbit is stable in $\Gamma$ when $\theta \in [0.33 \pi,  0.48 \pi]\cup[0.52 \pi,  0.78 \pi]$.\\\\
\textbf{(III):} when $\theta=\pi$, the orbit becomes the Broucke orbit.\cite{BR, Yan}
\begin{figure}
\includegraphics[scale=0.15]{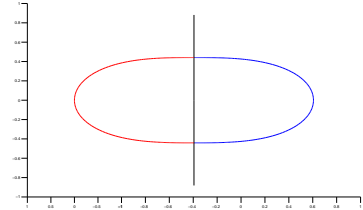}
\caption{\label{fig:broucke} Broucke orbit.}
\end{figure}
In each period of the Broucke orbit (Fig.~\ref{fig:broucke}), the middle body (body 2) moves up and down. When it reaches the highest or lowest point, the other two bodies collide.

The Broucke orbit  (Fig.~\ref{fig:broucke}) can be understood as a limit of the spatial isosceles orbits when $\theta \to \pi$. 
\begin{table}
\caption{\label{tab:table1}  This table lists the values of $a, b$ in the spatial isosceles periodic orbits corresponding to given rotation angle $\theta$.}
\begin{ruledtabular}
\begin{tabular}{ccc}
Vaule of $a$ in $Q(0)$ & Value of $b$ in $Q(1)$  \footnotemark[1]&  $\theta$ \\
\hline
0.8300  &  0.0062   &  0.96 $\pi$ \\
0.8335 & 0.0025  & 0.97 $\pi$ \\
 0.8337 &  2.51$\times 10^{-12}$ & 0.98 $\pi$ \\
  0.8320 &  4.21$\times 10^{-11}\footnotemark[2]  $ & 0.99 $\pi$ \\
\end{tabular}
\end{ruledtabular}
\footnotetext[1]{$Q(0)=\begin{bmatrix}
-a   &  0  & 0    \\
0  &  0  &   0   \\
a   & 0   &   0
\end{bmatrix}, \, Q(1)=   \begin{bmatrix}
-b   &  0  &  -c  \\
0  &  0   &   2c \\ 
b   &   0  &    -c
\end{bmatrix}\begin{bmatrix}
 \cos \theta   &  \sin \theta  &  0   \\
 - \sin \theta   &    \cos \theta    &   0 \\
0     &  0     &    1
\end{bmatrix}$.}
\footnotetext[2]{Due to the numerical error, the values of $a$ and $b$ at $\theta=0.99 \pi$ are not accurate enough.}
\end{table}
Actually in Table \ref{tab:table1}, as $\theta$ approaches $\pi$, the values of the parameter $b$ in $Q(1)$ of the corresponding spatial isosceles periodic orbits have a limit $0$. Note that in the matrix form \eqref{Q11} of $Q(1)$, bodies 1 and 3 experience a binary collision when $b=0$ and the action minimizer $\mathcal{P}_0$ becomes the Broucke orbit.\\\\
\textbf{(IV):} At $\theta=\pi/2$, an oscillated behavior occurs when perturbing the values of $a$, $v_y$ and $v_z$ in the initial condition matrix \eqref{pi2ini}. The initial condition of this orbit is
\begin{equation}\label{pi2ini}
\begin{bmatrix}
q_1 & \dot{q}_1 \\
q_2 & \dot{q}_2 \\
q_3 & \dot{q}_3
\end{bmatrix}=\begin{bmatrix}
-a     &     0     &     0        &     0      &         -v_y       &    -v_z           \\
 0               &    0       &       0      &      0      &         0        &      2 v_z \\
  a   &     0     &         0    &    0        &       v_y            &      -v_z         
  \end{bmatrix},
  \end{equation}
 where $a=0.74528$, $v_y= 0.7335$ and $v_z= 0.6733$.  The uniqueness of ODE guarantees that the perturbed orbit will stay in the symmetric subspace $\Gamma$ as in Eqn \eqref{gammadef}.
We illustrate the oscillated behavior by running the following perturbed initial condition (only varying $v_y$ from $0.7335$ to $0.7333$):
\begin{equation}\label{perturbed}
\begin{bmatrix}
 -0.74528     &     0     &     0        &     0      &         -0.7333       &    -0.6733           \\
 0               &    0       &       0      &      0      &         0        &      1.3466           \\
  0.74528   &     0     &         0    &    0        &       0.7333            &      -0. 6733         
  \end{bmatrix}.
  \end{equation} 
In this case, it keeps the periodic shape only for a few periods. The movements of bodies 1 and 3 then become chaotic. However, after a while the motion becomes periodic again,  and this periodic orbit looks like a rotation of the spatial isosceles orbit at $\pi/2$. The orbit keeps shifting between periodic motion and chaotic motion again and again. Similar behaviors happen when we slightly modify $a$ or $v_z$ in the initial condition matrix \eqref{pi2ini}. The orbit with perturbed initial condition in $\Gamma$ provides a concrete example of oscillation in the isosceles three-body problem which is far away from collision singularity.  

\begin{figure}
\begin{minipage}[t]{0.48\linewidth}
\centering
\includegraphics[width=1.8in]{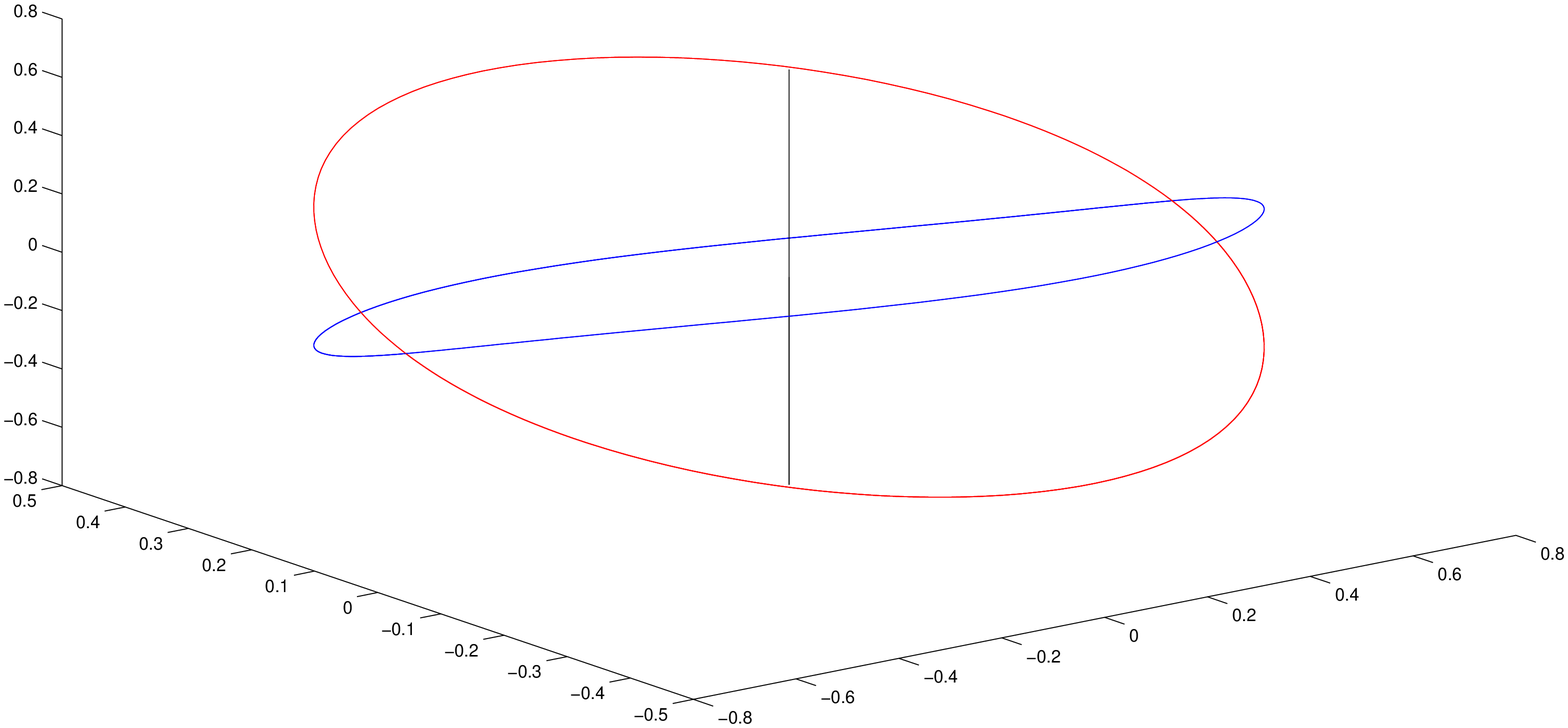}
\caption{Orbit for $t \in [0, 20].$}
\label{fig:a1}
\end{minipage}%
\begin{minipage}[t]{0.48\linewidth}
\centering
\includegraphics[width=1.8in]{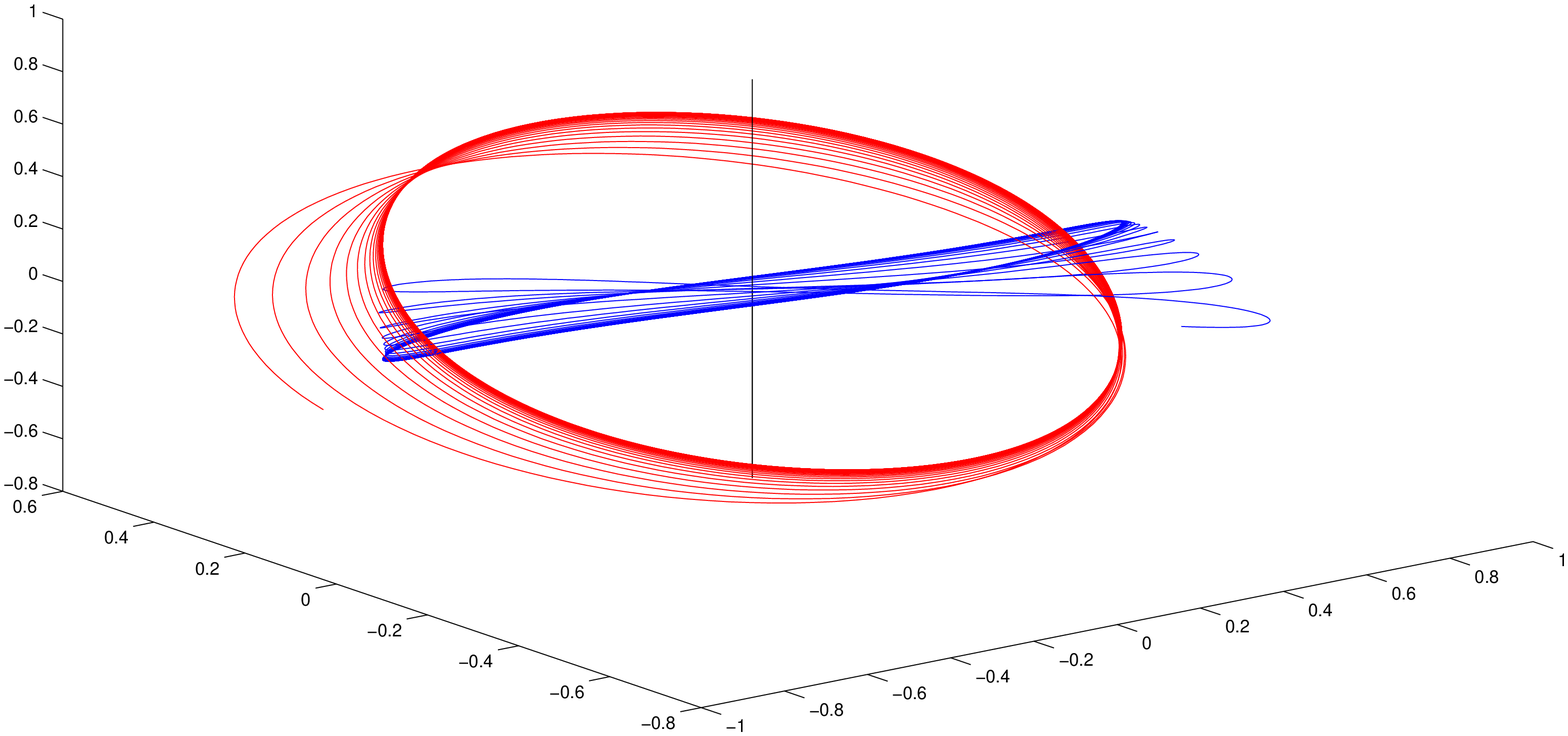}
\caption{ $t \in [0, 180].$}
\label{fig:a2}
\end{minipage}
\begin{minipage}[t]{0.48\linewidth}
\centering
\includegraphics[width=1.8in]{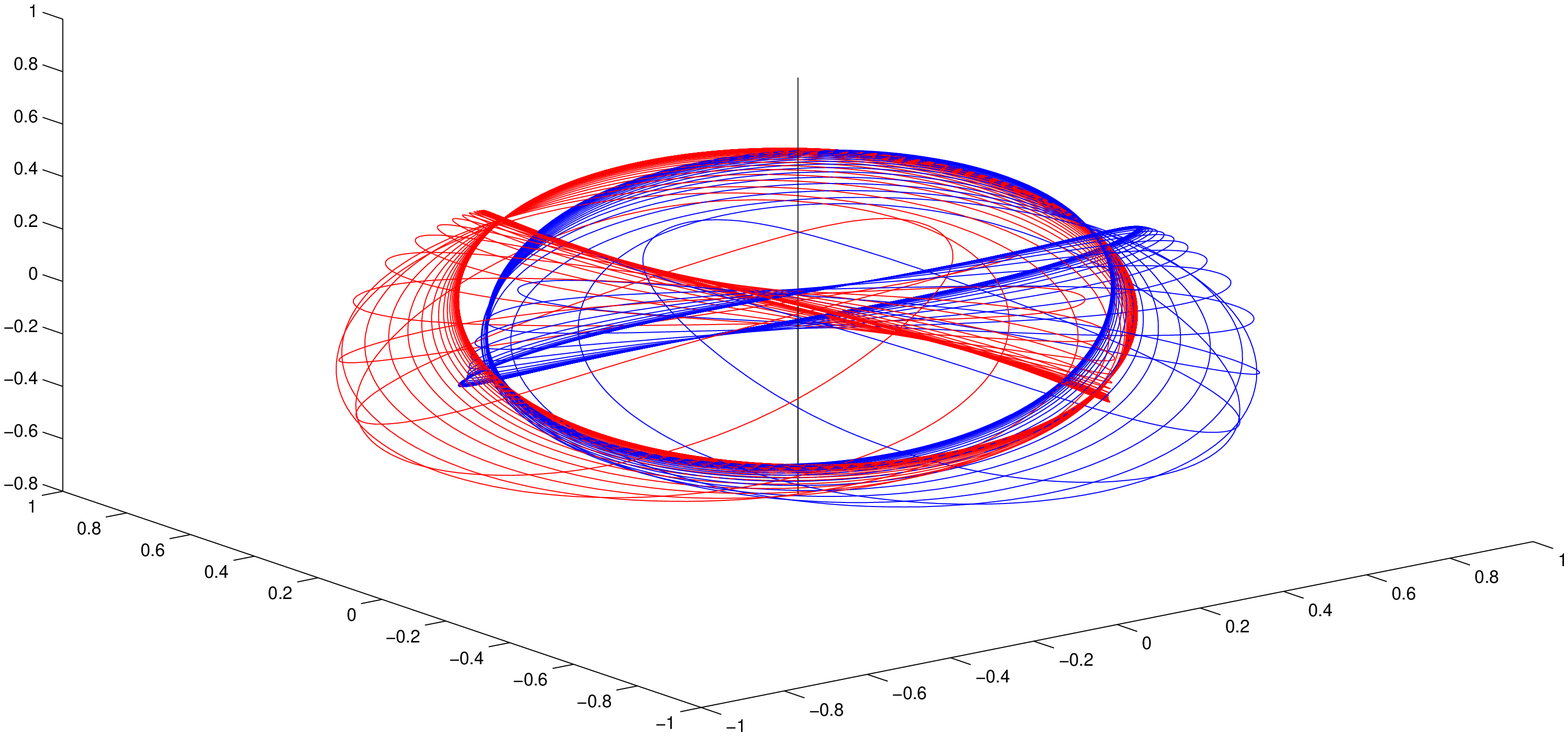}
\caption{  $t \in [0, 280].$}
\label{fig:a3}
\end{minipage}%
\begin{minipage}[t]{0.48\linewidth}
\centering
\includegraphics[width=1.8in]{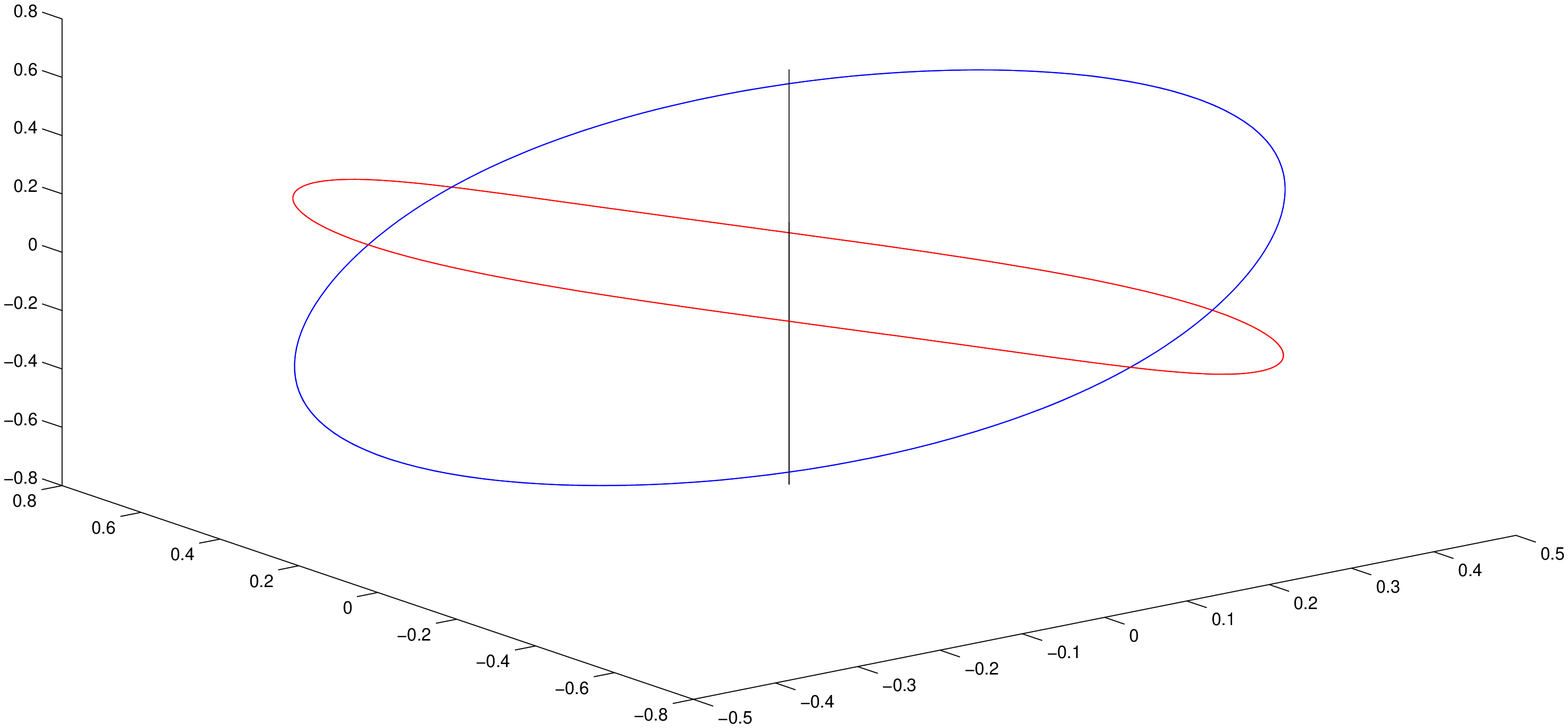}
\caption{  $t \in [280, 288].$}
\label{fig:a4}
\end{minipage}
\begin{minipage}[t]{0.48\linewidth}
\centering
\includegraphics[width=1.8in]{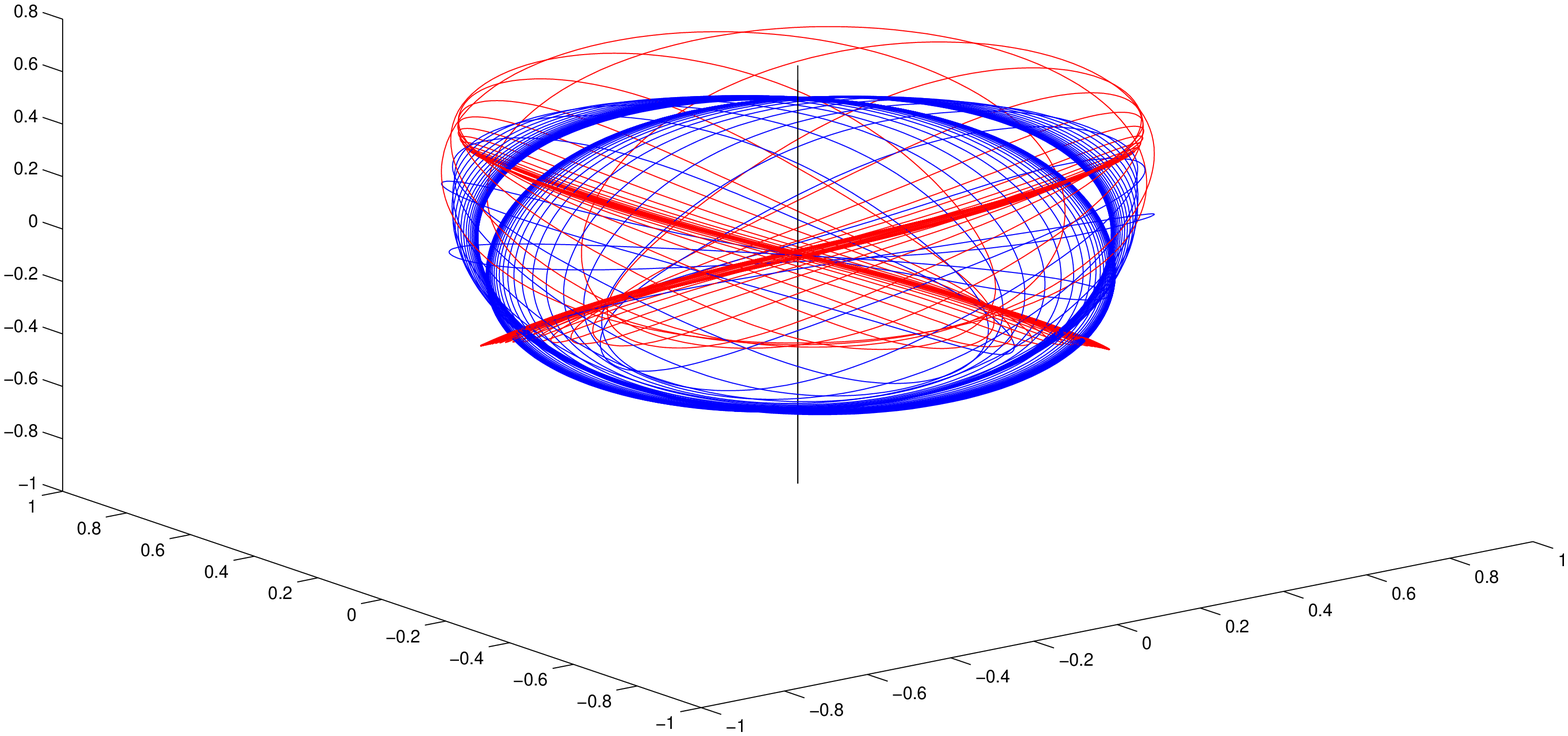}
\caption{  $t \in [280, 520].$}
\label{fig:b1}
\end{minipage}%
\begin{minipage}[t]{0.48\linewidth}
\centering
\includegraphics[width=1.8in]{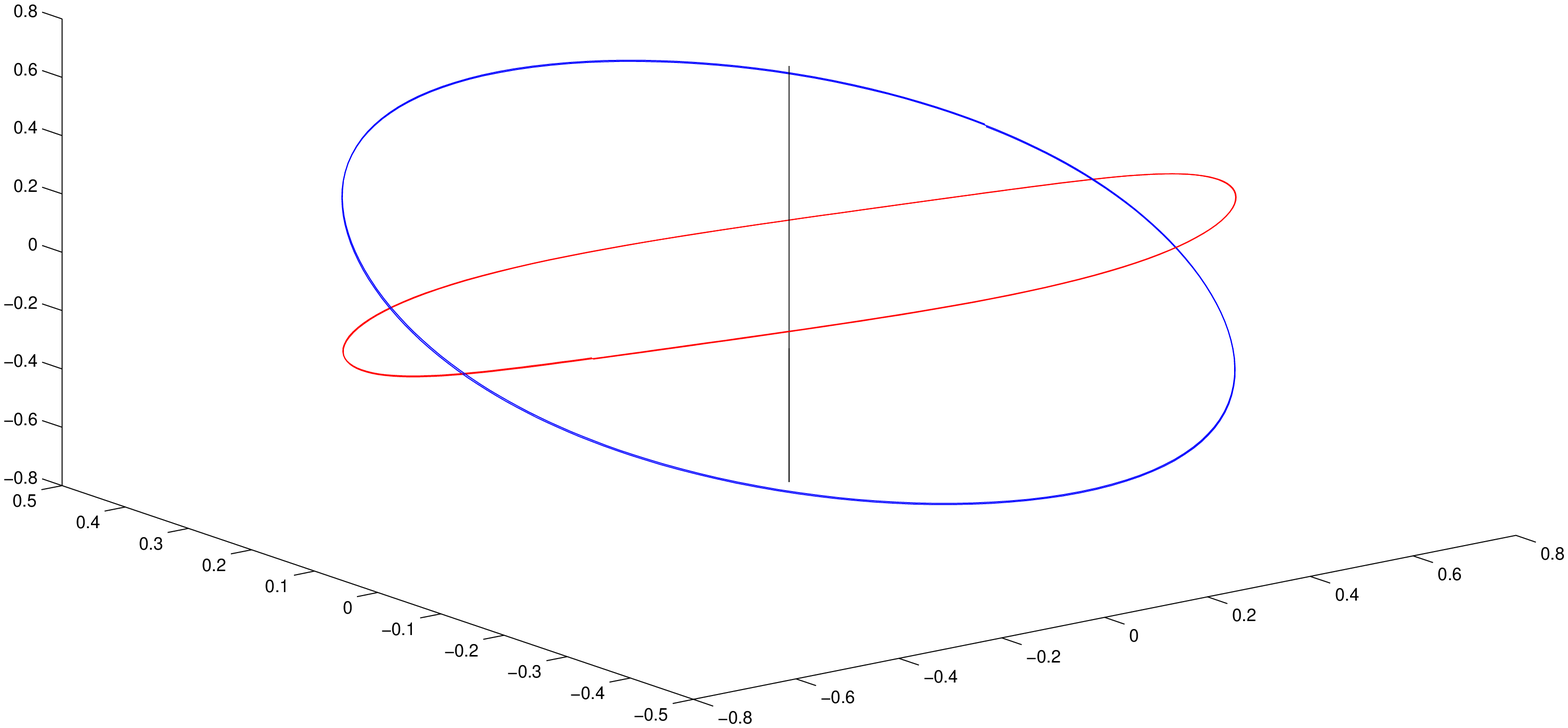}
\caption{  $t \in [520,528].$}
\label{fig:b2}
\end{minipage}
\begin{minipage}[t]{0.48\linewidth}
\centering
\includegraphics[width=1.8in]{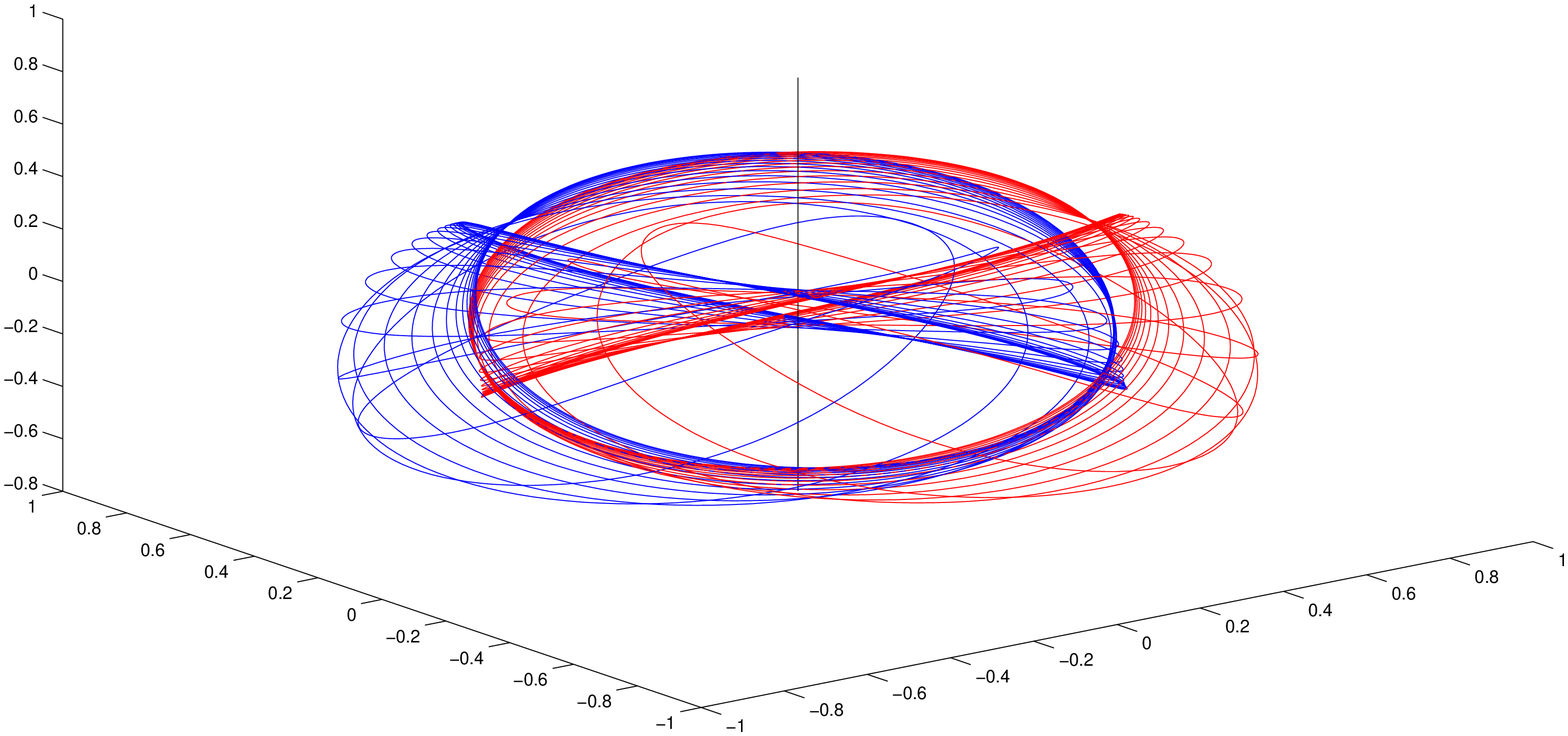}
\caption{ $t \in [520, 640].$}
\label{fig:b3}
\end{minipage}%
\begin{minipage}[t]{0.48\linewidth}
\centering
\includegraphics[width=1.8in]{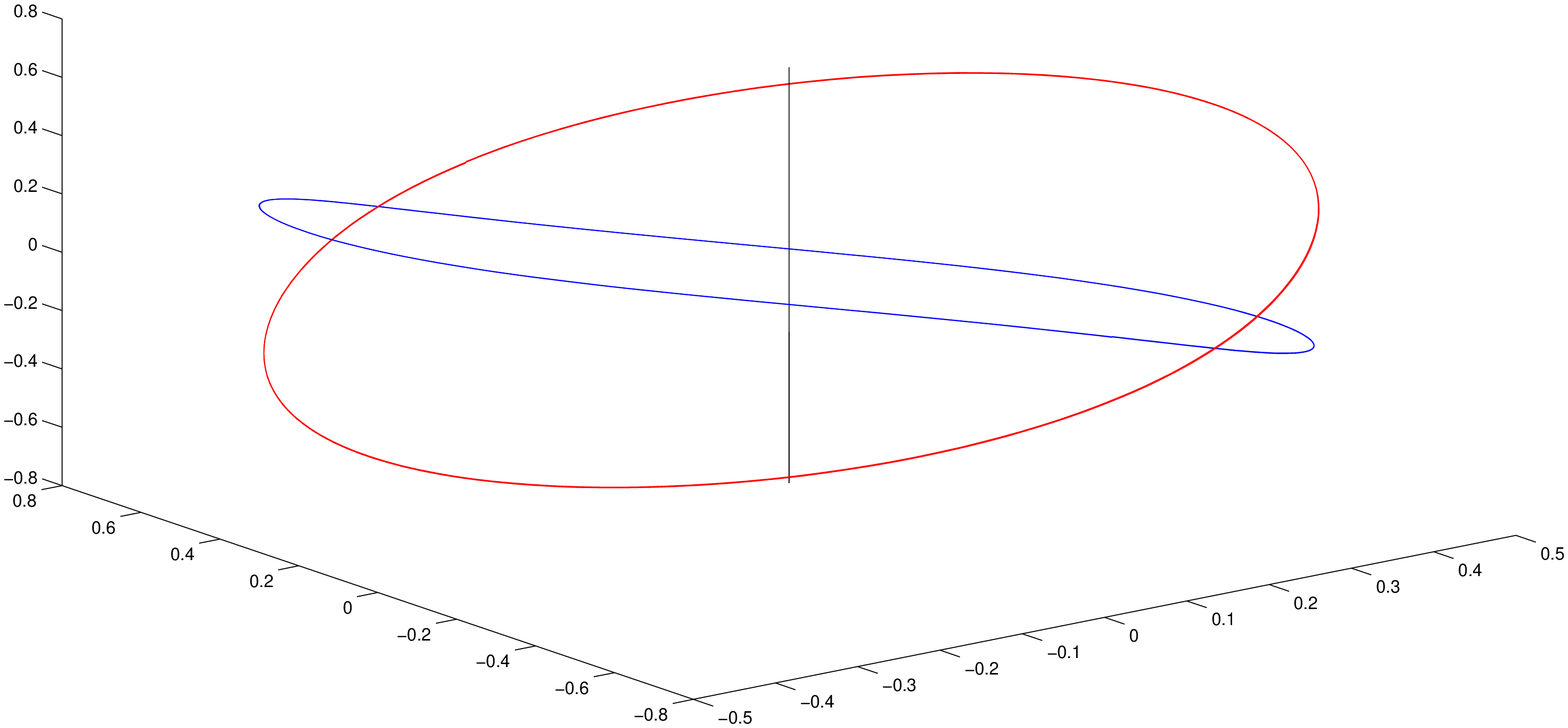}
\caption{ $t \in [640, 648].$}
\label{fig:b4}
\end{minipage}
\begin{minipage}[t]{0.48\linewidth}
\centering
\includegraphics[width=1.8in]{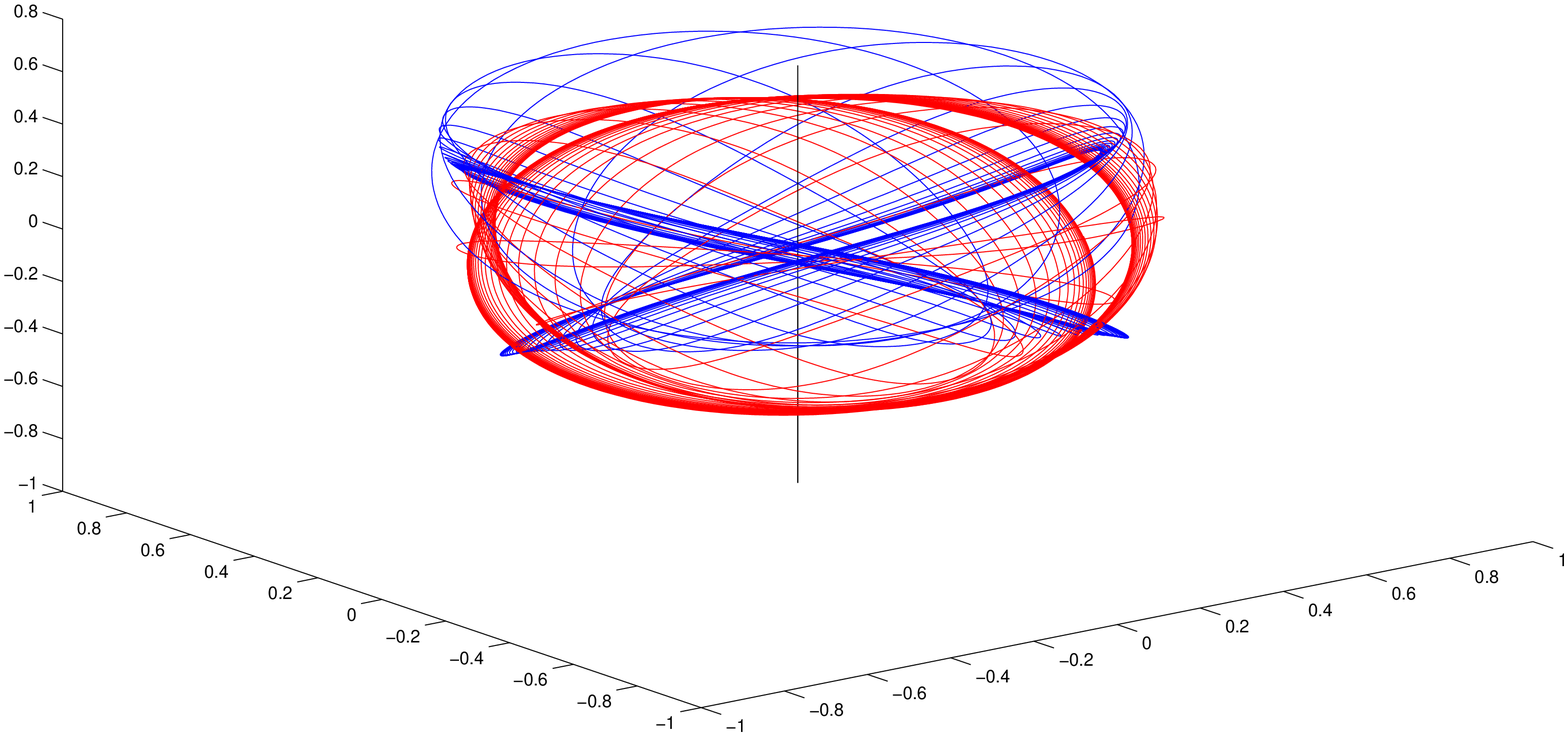}
\caption{  $t \in [640, 840].$}
\label{fig:b5}
\end{minipage}%
\begin{minipage}[t]{0.48\linewidth}
\centering
\includegraphics[width=1.8in]{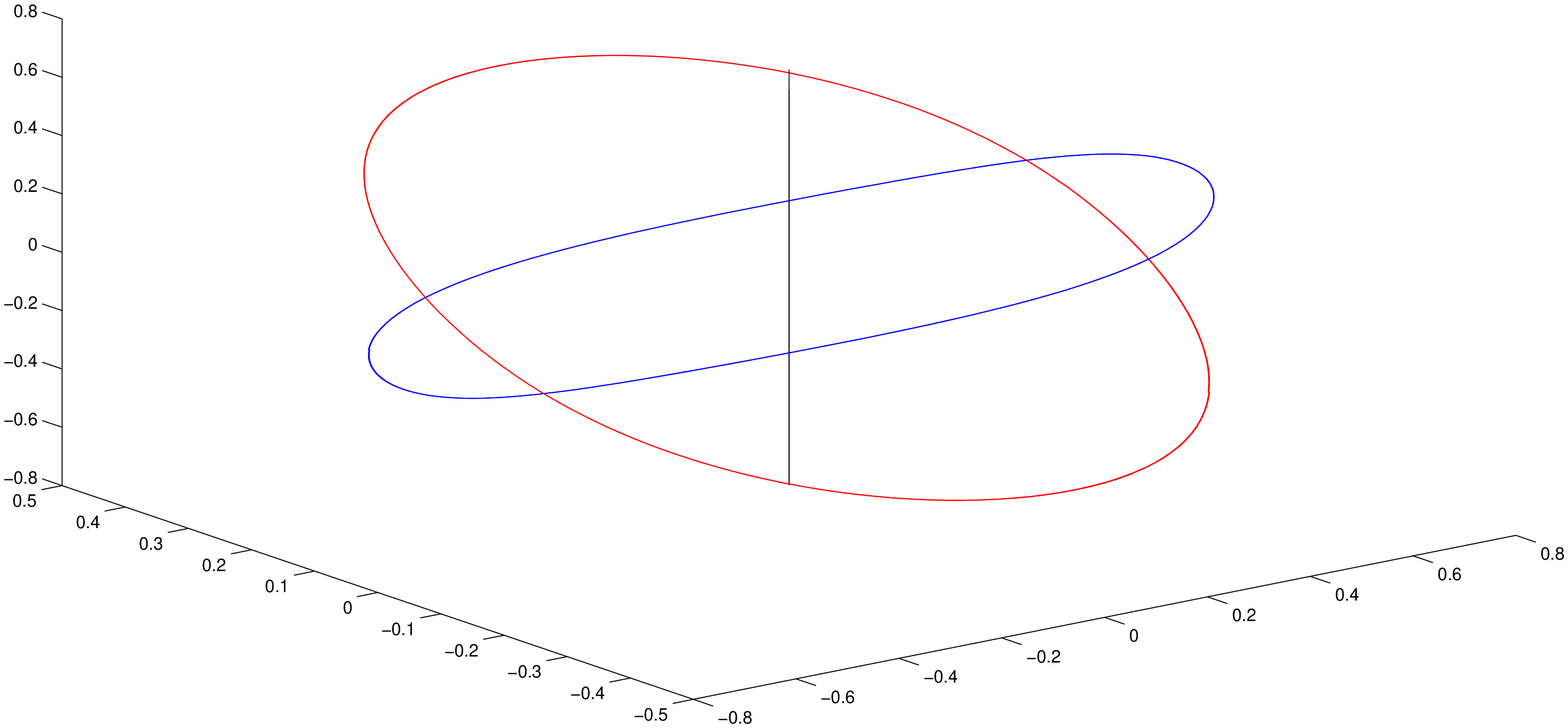}
\caption{ $t \in [840,848].$}
\label{fig:b6}
\end{minipage}
\end{figure}

In order to see the chaotic behavior clearly, we run the perturbed initial condition \eqref{perturbed} on our simulator for a long time so that one can see how the orbit changes from one periodic motion to another. For $t \in [0, 20]$, the motion of this orbit is shown in Fig.~\ref{fig:a1}.  As $t$ increases, the motion of this orbit becomes chaotic.  Fig.~\ref{fig:a2} shows the motion for $t \in [0, 180]$. It can be seen that bodies 1 and 3 move away from the original periodic orbit. We run this orbit a bit longer and see how would this motion end. Fig.~\ref{fig:a3} shows the motion for $t \in [0, 280]$. It seems that bodies 1 and 3 shift from one periodic orbit to another. A closer look at $t \in [280, 288]$ is shown in Fig.~\ref{fig:a4} and it confirms our guess. At $t=280$, the periodic motion in Fig.~\ref{fig:a4} is different from the motion in Fig.~\ref{fig:a1}. Actually, the oscillation does not stop at $t=280$. It keeps shifting from one periodic orbit to another. And the orbit stays bounded all the time. (Note that the time boundaries, $t=20, 180$ and $280$, etc., are not the exact time when the shape of orbit changes.)
After a long time run on the simulator, we find out that its motion oscillates between 4 periodic orbits. Fig.~\ref{fig:a1} to Fig.~\ref{fig:b6} shows how they shift from one to another. When $t \in [840, 848]$, the orbit in Fig.~\ref{fig:b6} is the same as Fig.~\ref{fig:a1}. (the orbit for $t \in [0,20]$.) 

\section{New periodic orbits in the N-body problem}
\begin{figure}[htb]
\includegraphics[scale=0.45]{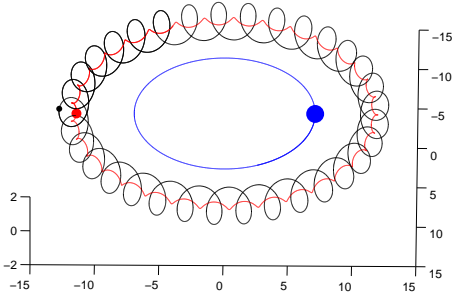}
\caption{\label{fig:3dsolar} 3D solar orbit with unequal masses}
\end{figure}

In the end, we present a few new orbits found by our variational method.  Fig.~\ref{fig:3dsolar} is a spatial periodic orbit in the three-body problem, which imitates the motion of the sun, earth and moon.
Its initial condition is
\begin{equation*}\label{initial3dsolar}
\begin{split}
& \begin{bmatrix}
q_1 & \dot{q}_1  & m_1 \\
q_2 & \dot{q}_2  &  m_2 \\
q_3 & \dot{q}_3  & m_3
\end{bmatrix}=\\
& \begin{bmatrix}
7.0467     &     0              &     0            &     0                        &         0.3458       &   0    &  10           \\
  -11.5217   &    0       &   -0.0452     &      0.0688         &         -0.5729           &     0.3399             &    5    \\
    -12.8589    &     0     &   0.2259    &     -0.3443           &       -0.5935               &      -1.6995           &      1
  \end{bmatrix},
  \end{split}
  \end{equation*}
where $q_i$ and $\dot{q}_i \, (i=1,2, 3)$ are all $1\times 3$ vectors.

\begin{figure}
\begin{minipage}[t]{0.48\linewidth}
\centering
\includegraphics[width=1.8in]{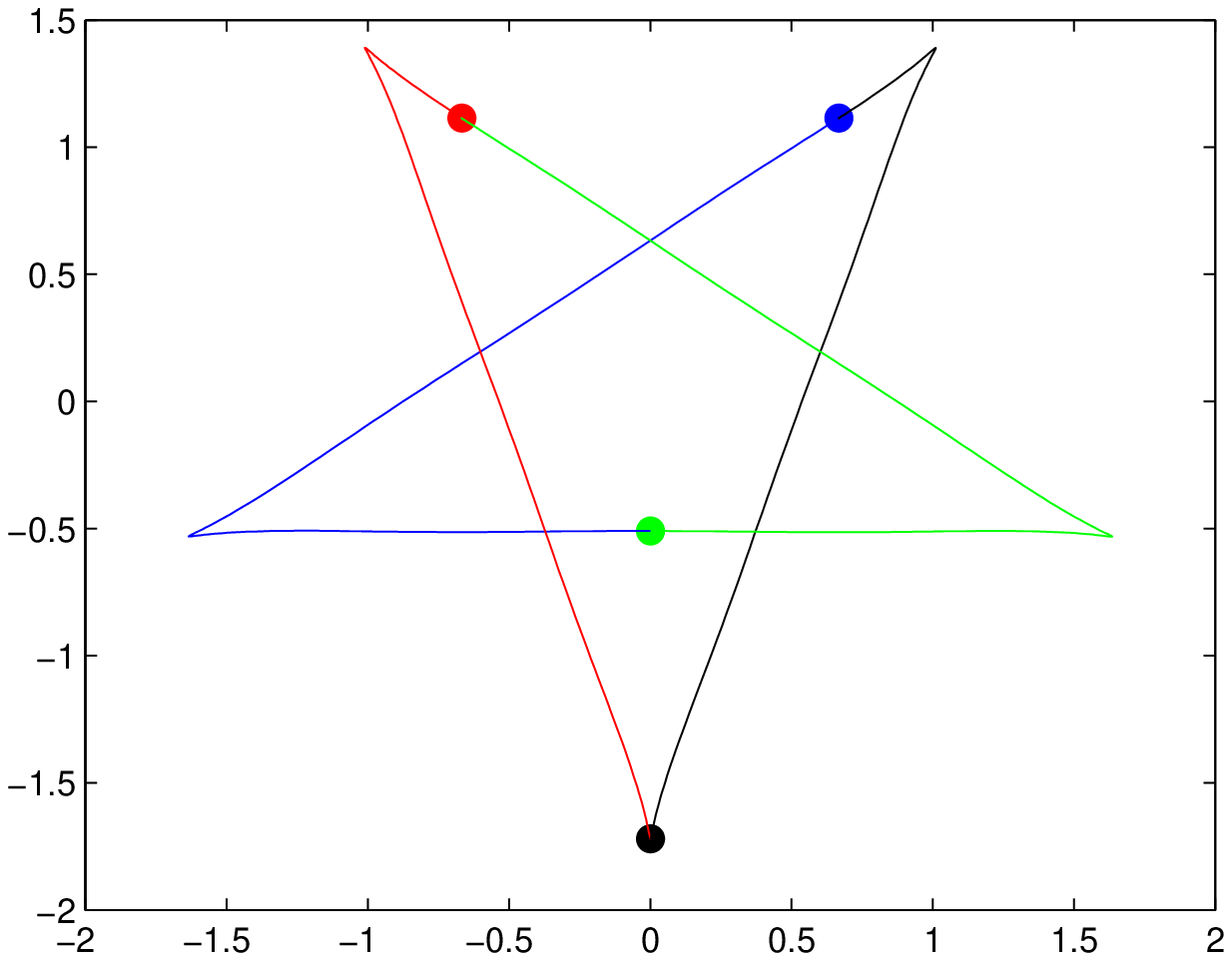}
\caption{4 body orbit 1}
\label{fig:d1}
\end{minipage}%
\begin{minipage}[t]{0.48\linewidth}
\centering
\includegraphics[width=1.8in]{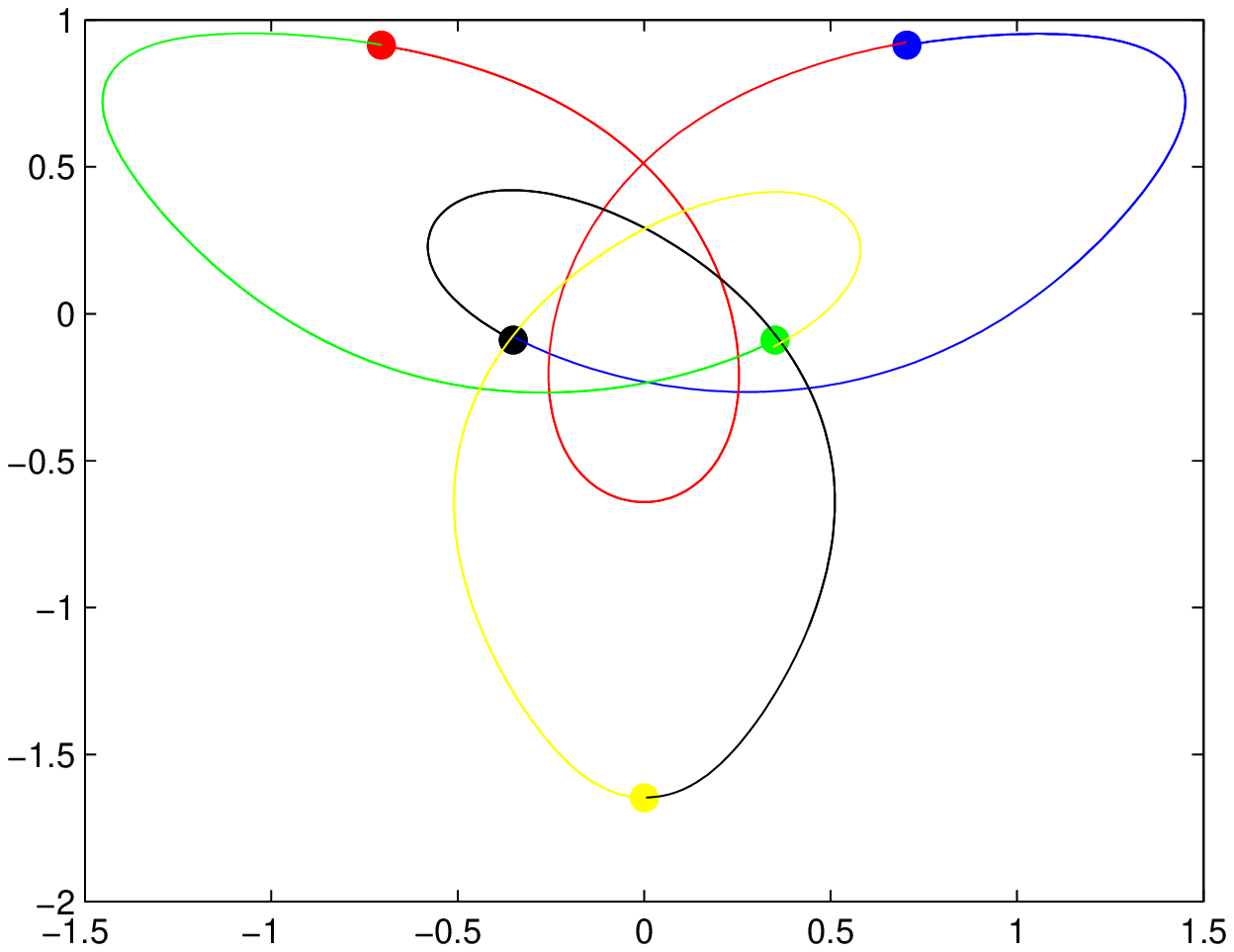}
\caption{5 body orbit 1}
\label{fig:d2}
\end{minipage}
\begin{minipage}[t]{0.48\linewidth}
\centering
\includegraphics[width=1.8in]{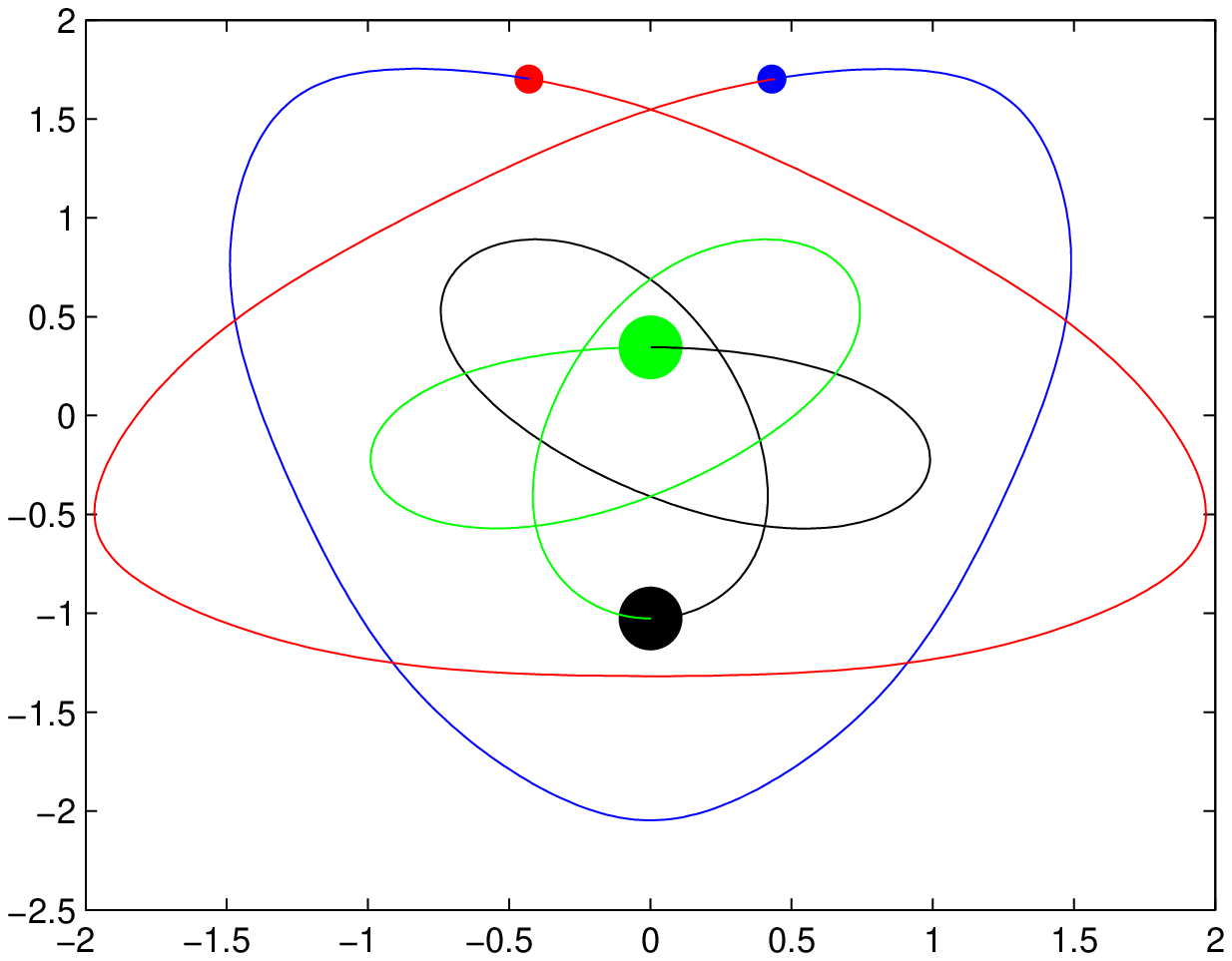}
\caption{4 body orbit 2}
\label{fig:d3}
\end{minipage}%
\begin{minipage}[t]{0.48\linewidth}
\centering
\includegraphics[width=1.8in]{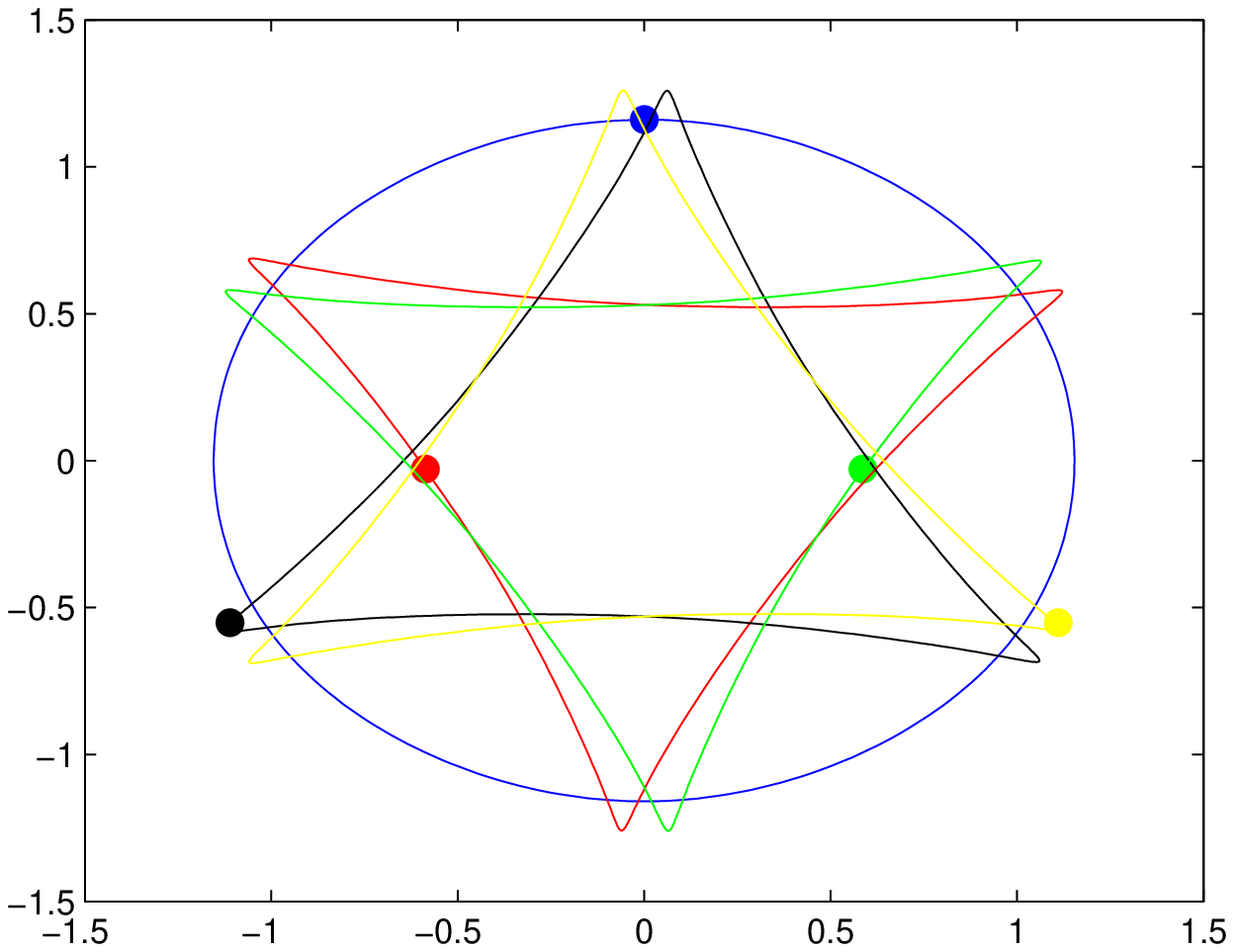}
\caption{5 body orbit 2}
\label{fig:d4}
\end{minipage}
\begin{minipage}[t]{0.48\linewidth}
\centering
\includegraphics[width=1.8in]{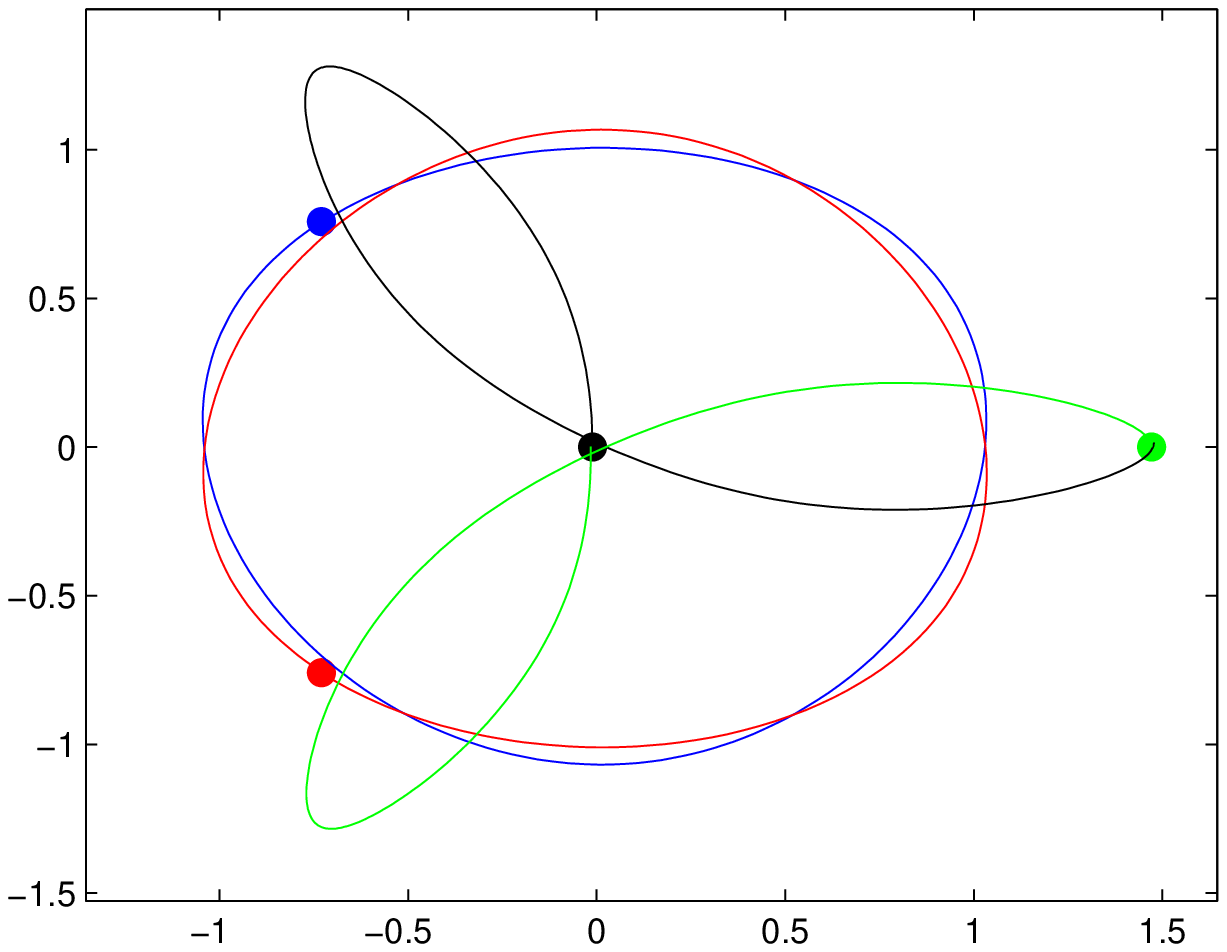}
\caption{4 body orbit 3}
\label{fig:d5}
\end{minipage}%
\begin{minipage}[t]{0.48\linewidth}
\centering
\includegraphics[width=1.8in]{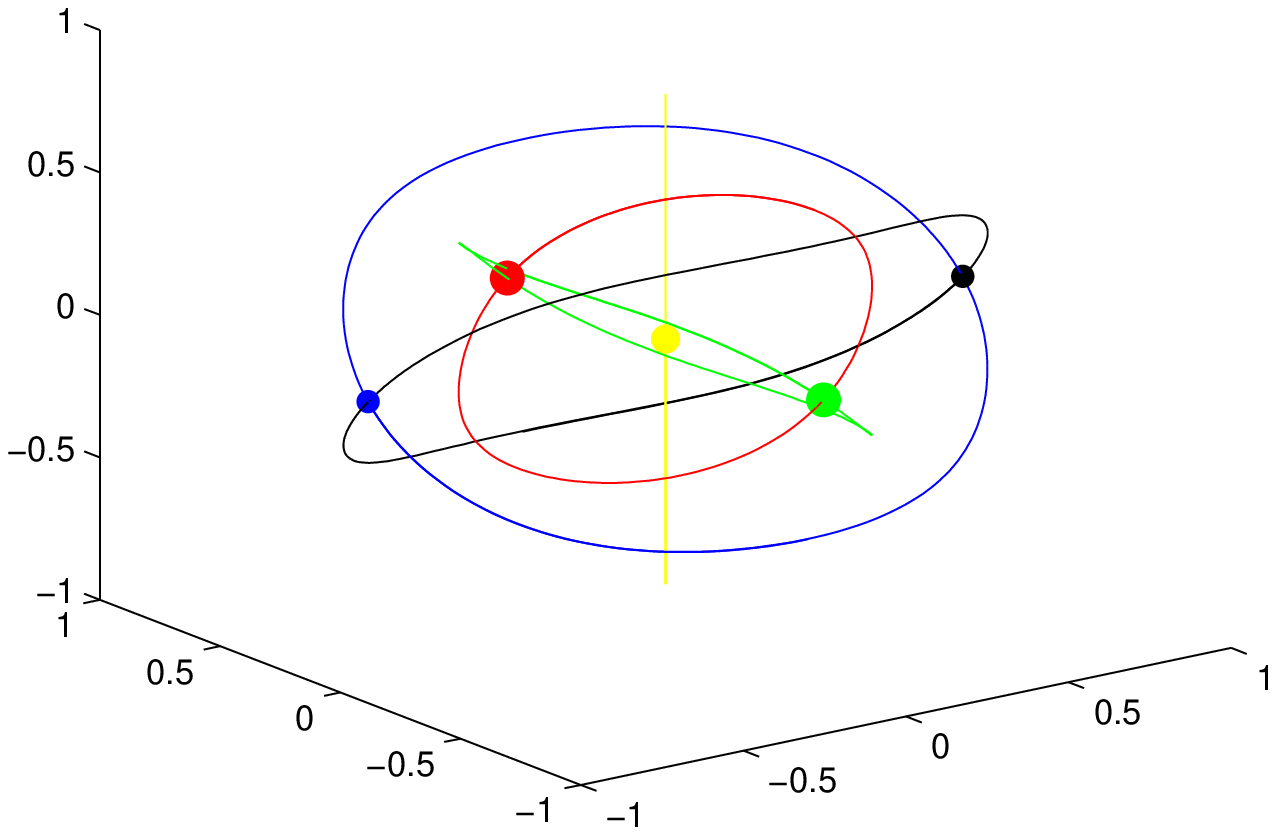}
\caption{5 body orbit 3}
\label{fig:d6}
\end{minipage}
\end{figure}

Fig.~\ref{fig:d1} to Fig.~\ref{fig:d6} are six periodic orbits in the four- or five-body problem.  Three of them (Fig.~\ref{fig:d1}, Fig.~\ref{fig:d3}, Fig.~\ref{fig:d5}) are four-body orbits and other three (Fig.~\ref{fig:d2}, Fig.~\ref{fig:d4}, Fig.~\ref{fig:d6}) are five-body orbits. In each figure,  the dots represent the starting positions in the orbit and the weight of masses are measured by their sizes. It is worth noting that Fig.~\ref{fig:d1} and Fig.~\ref{fig:d3} are apparently stable orbits.

For completeness, we list the initial condition matrices for these orbits. Each row of the matrices has the form $\begin{bmatrix}
q_i  &  \dot{q}_i & m_i
\end{bmatrix}$: the first 3 elements represent the position $q_i$, the next 3 elements are the velocity $\dot{q}_i$ and the last element is the mass $m_i$. 
\begin{equation*}
\begin{split}
& Fig.~\ref{fig:d1}: \\
& \begin{bmatrix}
0.6675    & 1.1148  &       0&   -0.6978 &   -0.5025  &       0 &   1\\
   -0.6675   & 1.1148      &   0   & -0.6978   & 0.5025        & 0   & 1\\
         0 &  -1.7197 &        0 &   0.0178 &  0 &       0 &    1\\
         0 &   -0.5099 &        0 &   1.3777  &  0 &         0 &    1
\end{bmatrix},
\end{split}
\end{equation*}
\begin{equation*}
\begin{split}
& Fig.~\ref{fig:d2}: \\
& \begin{bmatrix}
0.7045  &  0.9135 &      0 &   1.2833 &   0.2772 &        0  &  1 \\
   -0.7045  &  0.9135 &  0 &   1.2833 &  -0.2772  &       0 &   1 \\
   -0.3510 &  -0.0900 &  0 &  -1.0663  &  0.6946  &       0 &   1 \\
    0.3510   & -0.0900  & 0 &  -1.0663 &  -0.6946  &       0 &   1 \\
         0  & -1.6470 &  0 &   -0.4340  &  0 &         0   &   1 
\end{bmatrix},
\end{split}
\end{equation*}
\begin{equation*}
\begin{split}
& Fig.~\ref{fig:d3}: \\
& \begin{bmatrix}
0.4296    & 1.7008  &       0&   -2.1567 &   -0.5194  &       0 &   1\\
   -0.4296   & 1.7008      &   0   & -2.1567   & 0.5194        & 0   & 1\\
         0 &  -1.0259 &        0 &   -0.9477 &  0 &       0 &    5\\
         0 &   0.3456 &        0 &   1.8103  &  0 &         0 &    5
\end{bmatrix},
\end{split}
\end{equation*}
\begin{equation*}
\begin{split}
& Fig.~\ref{fig:d4}: \\
& \begin{bmatrix}
        0  &  1.1606  &       0   & 1.2298  &  0  &         0  &  1 \\
   -0.5863 &  -0.0290 &        0 &  -0.8492  &  1.5059  &       0  &  1 \\
   -1.1108 &  -0.5513   &      0  &  0.2343 &   0.2714   &      0 &   1 \\
    0.5863 &  -0.0290  &       0 &  -0.8492   & -1.5059  &       0 &   1 \\
1.1108  &  -0.5513 &        0 &   0.2343  &   -0.2714   &      0 &   1 
\end{bmatrix},
\end{split}
\end{equation*}
\begin{equation*}
\begin{split}
& Fig.~\ref{fig:d5}: \\
& \begin{bmatrix}
-0.7302  &  0.7585   &      0  & -0.8938 &  -0.7330  &       0 &   1\\
   -0.7302  &  -0.7585  &       0 &   0.8938 &  -0.7330  &       0   & 1 \\
   -0.0112   &      0   &      0  &  0  &    1.2701  &       0  &  1\\
    1.4717  &       0   &      0  &  0   &   0.1959   &      0  &   1
\end{bmatrix},
\end{split}
\end{equation*}
\begin{equation*}
\begin{split}
& Fig.~\ref{fig:d6}: \\
& \begin{bmatrix}
 -0.9145  &       0    &     0 &  0  &   -0.7385  &  -0.8869 &   0.25\\
         0   &   0.6579  &        0   &  0.6304  &  0  &    0.6415  &  1 \\
    0.9145   &      0  &       0 &   0  &    0.7385 &   -0.8869  &  0.25 \\
         0   &   -0.6579   &        0 &  -0.6304  &    0  &    0.6415 &   1  \\
         0    &     0   &      0   &  0  &    0  &   -1.6792 &    0.5  
\end{bmatrix}.
\end{split}
\end{equation*}

\section{Summary and discussion}

In this paper, we introduce a new variational method to investigate the spatial isosceles periodic orbit in the equal-mass three-body problem. There are basically two advantages of this method. First, this method does not require any symmetry constraint or equal-mass assumption, which allows a lot of flexibility. Whenever you have two special boundary configurations (2D or 3D) in mind, this method can help you identify if there is a nontrivial periodic orbit connecting them as a local action minimizer. And between two different configurations, there may exist several different periodic orbits as local minimizers. Many four-body periodic orbits \cite{OX, YOX} have been found recently by this method.  Second, this method provides a detailed variational property of the orbit, which is helpful in studying the variational existence and its linear stability. Actually, for the periodic orbits searched by this method, it presents a scheme of mathematical proofs. However, there are still shortcomings in our searching program. The main shortcoming is that the searching process is not efficient enough. It is because the minimizing functions we use are adopted from Matlab. To improve the searching program will be part of our projects. 
 
The motions of the spatial isosceles periodic orbits are studied in detail for the first time. They can be classified into four types. Particularly, when rotation angle $\theta=\pi/2$, a chaotic (or oscillated) behavior is discovered. It would be very interesting if one can show the existence of this chaotic behavior rigorously. We expect similar phenomenons when the masses become $[1, m,  1]$ in general. Several new 2D and 3D periodic orbits in N-body problem are presented in the end. Our next step is to search for possibly stable periodic orbits\cite{Van, OX} in the N-body problem with $N=4, \, 5$ or $6$ and classify them. 

\section*{Acknowledgements}
The authors, D. Yan and T. Ouyang, are equally contributed in this work. We sincerely thank Professor Yiming Long for his precious help and valuable discussions on these and related topics. D. Yan was supported by NSFC (No. 11101221). Part of this work was done while D. Yan was visiting Brigham Young University; he sincerely thanks the department of mathematics there for its help and support. 

\end{document}